\documentclass[%
 aip,
numerical,
 amsmath,amssymb,
 reprint,%
]{revtex4-1}

\usepackage{graphicx}
\usepackage{dcolumn}
\usepackage{bm}

\usepackage[utf8]{inputenc}
\usepackage[T1]{fontenc}
\usepackage{mathptmx}
\usepackage{etoolbox}

\makeatletter
\def\@email#1#2{%
 \endgroup
 \patchcmd{\titleblock@produce}
  {\frontmatter@RRAPformat}
  {\frontmatter@RRAPformat{\produce@RRAP{*#1\href{mailto:#2}{#2}}}\frontmatter@RRAPformat}
  {}{}
}%
\makeatother
\begin{document}


\title{Optical properties and exciton transfer between N-heterocyclic carbene iridium (III) complexes for blue light-emitting diode applications from first principles} 



\author{Irina V. Lebedeva}
\email{liv\_ira@hotmail.com (I. V. Lebedeva), j.jornet.somoza@gmail.com (J. Jornet-Somoza)}
\affiliation{Nano-Bio Spectroscopy Group and ETSF, Universidad del Pa\'is Vasco, CFM CSIC-UPV/EHU, 20018 San Sebasti\'an, Spain}

\author{Joaquim Jornet-Somoza}
\affiliation{Nano-Bio Spectroscopy Group and ETSF, Universidad del Pa\'is Vasco, CFM CSIC-UPV/EHU, 20018 San Sebasti\'an, Spain}
\affiliation{Max Planck Institute for the Structure and Dynamics of Matter and Center for Free-Electron Laser Science, Luruper Chaussee 149, 22761 Hamburg, Germany}

\date{\today}

\begin{abstract}
N-heterocyclic carbene (NHC) iridium (III) complexes are considered as promising candidates for blue emitters in organic light-emitting diodes. They can play the roles of the emitter as well as of electron and hole transporters in the same emission layer. We investigate optical transitions in such complexes with account of geometry and electronic structure changes upon excitation or charging and exciton transfer between the complexes from first principles. It is shown that excitation of NHC iridium complexes is accompanied by a large reorganization energy $\sim$0.7 eV and a significant loss in the oscillator strength, which should lead to low exciton diffusion. Calculations with account of spin-orbit coupling reveal a small singlet-triplet splitting $\sim$0.1 eV, whereas the oscillator strength for triplet excitations is found to be an order of magnitude smaller than for the singlet ones. The contributions of the F\"orster and Dexter mechanisms are analyzed via the explicit integration of transition densities. It is shown that for typical distances between emitter complexes in the emission layer, the contribution of the Dexter mechanism should be negligible compared to the F\"orster mechanism. At the same time, the ideal dipole approximation, although gives the correct order of the exciton coupling, fails to reproduce the result taking into account spatial distribution of the transition density. For charged NHC complexes, we find a number of optical transitions close to the emission peak of the blue emitter with high exciton transfer rates that can be responsible for exciton-polaron quenching. The nature of these transitions is analyzed.
\end{abstract}


\maketitle 

\section{Introduction}
Organic semiconductors \cite{Chiang1977} are strong competitors of traditional inorganic semiconductors winning in low-cost production, easy processability, flexibility and chemical tunability \cite{Yin2016,Parashkov2005,Boudreault2011, Bundgaard2007,Shirota2007,Huang2019, Kang2021,Yu2018,Dubey2019,Jou2018,Xiao2022}. Organic light emitting diodes (OLEDs) have already become a part of our everyday life as the principal ingredient of active matrix displays and are now routinely produced at the industrial scale \cite{Forrest2004, Hung2002, Zou2020,Chen2018, Kim2023, Graydon2023,Siddiqui2023,Swayamprabha2021}. While decisive steps have been also taken for the use of OLEDs in lighting applications, a number of problems including white light generation \cite{Sun2006,Schwartz2007,Reineke2009}, roll-off of quantum efficiency at high current densities \cite{Baldo2000a,  Zang2008, Song2010, Wehrmeister2015, Giebink2008}, efficient light outcoupling \cite{Lupton2000,Melpignano2006,Will2018} and long-term stability \cite{Seifert2013,Niu2017,Swayamprabha2021,Trindade2023} still have to be solved.

Interaction of  electrons and holes in OLEDs leads to formation of both singlet and triplet excitons. Spin statistics provides that only 25\% of energy can be extracted from singlet excitons through the fluorescence process and this is the limit of the internal quantum efficiency for OLEDs based on purely organic fluorescent emitters \cite{Baldo1999, Reufer2005}. It is possible, however, to take advantage of 75\% of the energy left in triplet excitons through the phosphorescence processes mediated by spin-orbit coupling \cite{Baldo1998, Adachi2001}. The use of organometallic phosphors embedded as guest molecules in the emission layer of OLEDs has allowed to approach the nearly 100\% internal efficiency, high luminance and long device lifetimes in the green \cite{Adachi2000,Kang2021} and red  \cite{Adachi2001a} spectral ranges. Nevertheless, the development of stable and efficient phosphorescent blue emitters remains a challenge \cite{Erk2006, Rothmann2011, Zhang2014, Niu2017, Trindade2023, Swayamprabha2021, Izawa2023, Heimel2018, Kim2022, Yan2023,Siddiqui2023}. 

Although OLEDs based on phosphorescent complexes have the high internal quantum efficiency, the external efficiency of such systems declines strongly with increasing the current \cite{Reineke2007, Baldo2000a,Song2010, Zang2008,Wehrmeister2015,Giebink2008}. The reason is that triplet excitons have long radiative lifetimes in the microsecond range \cite{Reineke2007, Baldo2000, Li2005, Tsuboyama2003, Lamansky2001,Giebink2008,Wehrmeister2015,Haneder_tesis, Haneder2008, Mesta2016, Eersel2014, Eersel2016, Coehoorn2017, Coehoorn2017a, Vries2019, Vries2020, Ligthart2021, Ligthart2018, Mori2014} (as compared to nanoseconds for fluorescent dyes \cite{Giebink2008, Heimel2018, Paterson2019}) and exciton quenching through interactions with polarons \cite{Reineke2007, Kalinowski2006, Song2010, Zang2008, Siboni2012, Mesta2016, Eersel2014, Eersel2016, Coehoorn2017, Wehrmeister2015, Ligthart2021, Zhao2023}, other excitons \cite{Reineke2007, Baldo2000a, Staroske2007, Song2010,  Mesta2016, Eersel2014, Eersel2016, Coehoorn2017, Coehoorn2017a, Wehrmeister2015, Ligthart2018, Coehoorn2019, Niyonkuru2022} and electric field \cite{Kalinowski2006} become important (see also review \cite{Murawski2013}). In addition to exciton loss, energy transfer to excited or charged species can provoke their dissociation and chemical degradation of the device \cite{Scholz2008, Moraes2011,Jurow2014,Scholz2015,Trindade2023}. To improve the OLED efficiency it is, therefore, indispensable to understand the nature of optical transitions in phosphorescent complexes and microscopic mechanisms of exciton transfer. In the present paper, they are analyzed using \textit{ab initio} calculations.

\begin{figure}
 \centering
 \includegraphics[width=\columnwidth]{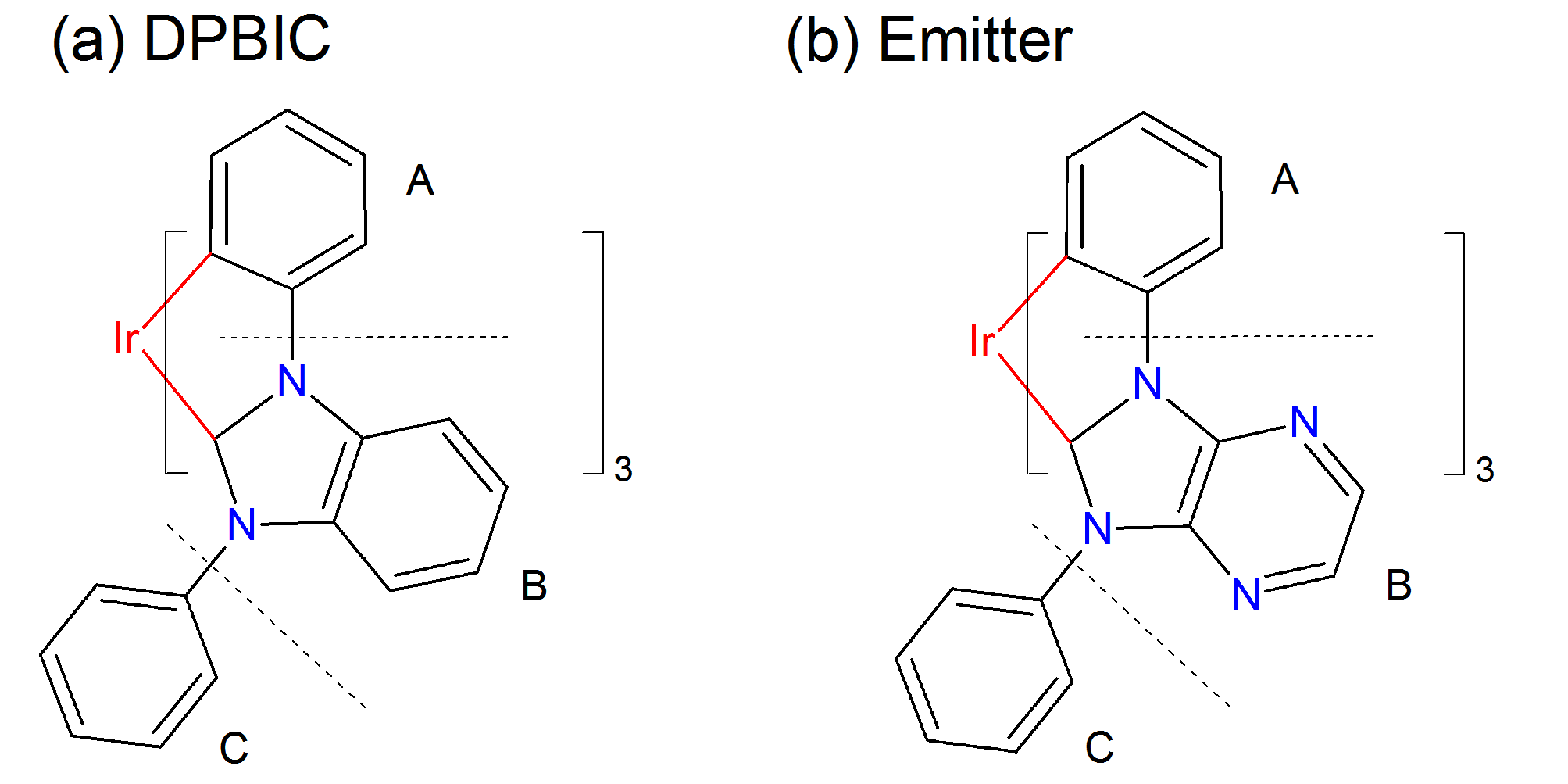}
 \caption{Chemical structure of (a) DPBIC (tris[(3-phenyl-1H-benzimidazol-1-yl-2(3H)-ylidene)-1,2-phenylene]-iridium) and (b) NHC blue emitter (tris[(3-phenyl-1H-imidazo[4,5-b]pyrazine-1-yl-2(3H)-ylidene)-1,2-phenylene]-iridium). For convenience, the fragments of the ligands are denoted as A (phenylene group), B and C (phenyl group).}
 \label{fig:fragments} 
\end{figure}

A low non-radiative loss of excitons and high efficiency can be expected for emitters with a short phosphorescence lifetime, which corresponds to a large spin-orbit coupling and a small singlet-triplet splitting \cite{Haneder_tesis, Haneder2008}. For this reason, cyclometalated octahedral iridium (III) complexes have been actively studied in the recent years \cite{Li2005, Tsuboyama2003, Lamansky2001, Adachi2001, Adachi2001a, Adachi2000, Baldo1999, Vries2019, Vries2020, Ligthart2021, Ligthart2018, Mori2014, Coehoorn2019,Yan2023,Kang2021,Yu2018,Trindade2023} including N-heterocyclic carbene (NHC) iridium (III) complexes\cite{Sajoto2005,Schildknecht2005,Erk2006,Haneder_tesis, Haneder2008, Heimel2018, Paterson2019, Yan2023}. A high radiative decay rate and small singlet-triplet splitting\cite{Haneder_tesis, Haneder2008} is achieved for tris[(3-phenyl-1H-benzimidazol-1-yl-2(3H)-ylidene)-1,2-phenylene]-iridium (DPBIC, Fig.  \ref{fig:fragments}a), which emits in the ultraviolet region. Substitution of two carbon atoms by nitrogen in each ligand (Fig.  \ref{fig:fragments}b) shifts the emission to the blue range, while keeping the radiative decay rate large \cite{Niu2017,Heimel2018,Paterson2019}. This complex allows to achieve longer device lifetimes as compared to alternative blue phosphorescent emitters \cite{Niu2017,Heimel2018,Paterson2019}.  Following the OLED architecture proposed in Ref. \onlinecite{Niu2017}, we assume that the emission layer is composed of the wide band gap matrix doped with the blue phosphorescent NHC
dye, which also  plays the role of the electron transporter, and co-doped with DPBIC, which plays the role of the hole transporter, to guarantee the charge balance.
Since there is no exciton transfer to the wide band gap matrix (as confirmed in our calculations), we investigate optical transitions in NHC dye and DPBIC complexes with account of geometry and electronic structure changes upon excitation and charging as well as exciton transfer between the complexes. This allows us to make qualitative predictions regarding the character of exciton dynamics in the emission layer and to get an insight into exciton-polaron quenching.

The exciton transport in OLEDs is generally described using two major non-radiative energy transfer mechanisms. In the F\"orster model  \cite{Forster}, the exciton transfer takes place via dipole-dipole interaction of the donor and acceptor and can be considered as the emission of a virtual photon by the donor followed by its instant re-absorption by the acceptor. The exciton coupling in the F\"orster model thus decays as $R_\mathrm{DA}^{-3}$ with the distance $R_\mathrm{DA}$ between the donor and acceptor. The Dexter coupling  \cite{Dexter} is based on the correlated transfer of two electrons due to exchange interaction and requires the spatial overlap of the donor and acceptor orbitals. It is, therefore, a short-ranged mechanism and the corresponding exciton coupling decreases exponentially with the donor-acceptor separation. While the combined spin is conserved in the Dexter mechanism making possible both triplet-triplet and singlet-singlet transport, the F\"orster mechanism applies only to optically-allowed transitions and triplet transport is prohibited. In materials with a large spin-orbit coupling, however, mixing between the singlet and triplet states provides that triplet transitions become optically allowed.  Although their oscillator strength is still low compared to singlet ones, the triplet transport also becomes possible via the F\"orster mechanism\cite{Baldo2000}. In the present paper, we analyze the contributions of these mechanisms for the exciton transfer between NHC dye complexes in the framework of time-dependent density functional theory (TDDFT). To model the dynamics of triplet excitons, the spin-orbit coupling is fully taken into account through magnetic structure calculations. 

The paper is organized as follows. First the methodology for calculation of exciton coupling and transfer rates is briefly reviewed. Then we consider singlet excitations in neutral and charged DPBIC and NHC blue emitter complexes and triplet excitations of the emitter. After that, contributions of the F\"orster and Dexter mechanisms to the exciton transfer in a dimer of NHC blue emitter complexes are analyzed. Transfer rates for singlet and triplet transport are estimated. Finally conclusions are summarized.

\section{Methods}
\subsection{Exciton transfer mechanisms}
We consider the energy transfer process
\begin{equation} \label{DA}
\mathrm{D}^*+\mathrm{A} \to \mathrm{A}^* + \mathrm{D},
\end{equation}
where $\mathrm{D}$ ($\mathrm{A}$) is donor (acceptor) and the asterisk denotes the excited state. In the system we study, the donor is the NHC blue emitter and the acceptor can be another blue emitter complex, DPBIC or charged species (positively or negatively charged blue emitter or positively charged DPBIC \cite{Niu2017}).

Based on the Fermi golden rule, the rate of this process can be written as \cite{Curutchet2017,Stehr2016,Hofmann2010,Forster,Malic2014}
\begin{equation} \label{k_F}
k=\frac{2\pi}{\hbar} |V_\mathrm{DA}|^2 J_\mathrm{DA},
\end{equation}
where $V_\mathrm{DA}=\langle \Psi_\mathrm{DA^*} |V_\mathrm{C}|\Psi_\mathrm{D^*A}\rangle$ is the electronic coupling between the initial and final states of the system described by wavefunctions $\Psi_\mathrm{D^*A}$ and $\Psi_\mathrm{DA^*}$, respectively, via the Coulomb interaction $V_\mathrm{C}$, $J_\mathrm{DA}$ is the spectral overlap between the normalized spectra of donor emission and acceptor absorption and $\hbar$ is the Planck constant. 

Assuming that the interaction between the donor and acceptor is very weak, the initial and final states can be written through the product states of the donor and acceptor as  \cite{Hofmann2010}
\begin{equation} \label{W1}
|\Psi_\mathrm{D^*A} \rangle= \frac{1}{\sqrt{2}}\left(|\Psi_\mathrm{D^*}\rangle|\Psi_\mathrm{A}\rangle-|\Psi_\mathrm{A}\rangle|\Psi_\mathrm{D^*}\rangle\right)
\end{equation}
and 
\begin{equation} \label{W2}
|\Psi_\mathrm{DA^*}\rangle = \frac{1}{\sqrt{2}}\left(|\Psi_\mathrm{D}\rangle|\Psi_\mathrm{A^*}\rangle-|\Psi_\mathrm{A^*}\rangle|\Psi_\mathrm{D}\rangle\right).
\end{equation}

Correspondingly the coupling is split into two components $V_\mathrm{DA}=V_\mathrm{F}+V_\mathrm{D}$ \cite{Curutchet2017,Stehr2016,Hofmann2010, Malic2014}, where
\begin{equation} \label{FRET}
V_\mathrm{F}=e^2\int\int \mathrm{d}\mathbf{r}\mathrm{d}\mathbf{r}'\frac{\rho^{\mathrm{T}*}_{\mathrm{A}} (\mathbf{r}) \rho^\mathrm{T}_{\mathrm{D}} (\mathbf{r}')}{|\mathbf{r}-\mathbf{r}'|}
\end{equation}
corresponds to the F\"orster energy transfer \cite{Forster} and 
\begin{equation} \label{Dexter}
V_\mathrm{D}=-e^2\sum_{\sigma,\sigma'}\int\int \mathrm{d}\mathbf{r}\mathrm{d}\mathbf{r}'\frac{\gamma^{\mathrm{T}*}_{\mathrm{A},\sigma\sigma'} (\mathbf{r},\mathbf{r}') \gamma^\mathrm{T}_{\mathrm{D},\sigma\sigma'} (\mathbf{r},\mathbf{r}')}{|\mathbf{r}-\mathbf{r}'|}
\end{equation}
describes the Dexter energy transfer \cite{Dexter}. Here $\mathbf{r}$ and $\mathbf{r}'$ are real-space vectors,  $\sigma$ and $\sigma'$ correspond to spin up ($\uparrow$) or spin down ($\downarrow$) and $e$ is the elementary charge. We also introduce here the notations for the transition density matrix related to the excited and ground-state wavefunctions $\Psi_\mathrm{M^*}$ and $\Psi_\mathrm{M}$ of a molecule $M$, respectively, as
\begin{equation} \label{gamma}
\begin{split}
\gamma^\mathrm{T}_{\sigma\sigma'}& (\mathbf{r},\mathbf{r}')=N\sum_{\sigma_2,\sigma_3,...,\sigma_N}\int...\int \mathrm{d}\mathbf{r}_2\mathrm{d}\mathbf{r}_3...\mathrm{d}\mathbf{r}_N \\ & \Psi^*_\mathrm{M^*}(\sigma\mathbf{r},\sigma_2\mathbf{r}_2,...,\sigma_N\mathbf{r}_N) \Psi_\mathrm{M}(\sigma'\mathbf{r}',\sigma_2\mathbf{r}_2,...,\sigma_N\mathbf{r}_N), 
\end{split}
\end{equation}
and transition density $\rho^\mathrm{T}(\mathbf{r})=\sum_\sigma\gamma_{\sigma\sigma}^\mathrm{T} (\mathbf{r},\mathbf{r})$. As reflected in Eqs. (\ref{FRET}) and (\ref{Dexter}), only optically allowed transitions are involved in the F\"orster mechanism, while spin flip is possible in the Dexter mechanism. It is also seen from Eq. (\ref{Dexter}) that the Dexter mechanism depends on the overlap of the donor and acceptor states and, therefore, falls exponentially upon increasing the distance between them. 

Assuming that the size of the donor and acceptor molecules is small compared to the distance $R=|\mathbf{R}|$ between their centers of mass, Eq. (\ref{FRET}) can be expanded to powers of $R^{-1}$ (Ref. \onlinecite{Hofmann2010}) and the first term of this expansion is given by 
\begin{equation} \label{IDA}
V_\mathrm{F}^\mathrm{IDA}=\kappa_{\mathrm{AD}} \frac{d^{\mathrm{T}*}_{\mathrm{A}} d^{\mathrm{T}}_{\mathrm{D}}}{R^3}.
\end{equation}
Here
\begin{equation} \label{dipole}
\mathbf{d}^{\mathrm{T}} = -e\int  \mathrm{d}\mathbf{r}  \rho^\mathrm{T}(\mathbf{r}) \mathbf{r} 
\end{equation}
is the transition dipole moment, $d^{\mathrm{T}}=|\mathbf{d}^{\mathrm{T}}|$ and
\begin{equation} \label{kappa}
\kappa_{\mathrm{AD}}=\mathbf{e}_\mathrm{A}\mathbf{e}_\mathrm{D}-3\frac{(\mathbf{e}_\mathrm{A},\mathbf{R})(\mathbf{e}_\mathrm{D},\mathbf{R})}{R^2}
\end{equation}
is the factor describing the relative orientation of unit dipole moments $\mathbf{e} = \mathbf{d}^{\mathrm{T}}/d^{\mathrm{T}}$. Eq. (\ref{IDA}) simply corresponds to the dipole-dipole coupling of the transition dipole moments of the donor and acceptor and is referred to below as the ideal dipole approximation (IDA).

\subsection{Exciton coupling from TDDFT}
The large size of the iridium complexes combined with the necessity of taking into account spin-orbit coupling makes \textit{ab initio} calculations of their optical properties computationally expensive. Therefore, in the present paper we use time-dependent density functional theory (TDDFT) \cite{Marques2012,Ullrich2012,Castro2004,Marques2004,Marques2006}, which provides a sufficient accuracy at a moderate computational cost.

In TDDFT \cite{Marques2012,Ullrich2012}, the time-dependent Kohn-Sham equations \cite{Runge1984} are considered:
\begin{equation} \label{eq1}
i\hbar\frac{\partial}{\partial t} \phi_{m\sigma}\left(\mathbf{r}, t\right) = H_{KS,\sigma\sigma'}\left(\mathbf{r},t\right) \phi_{m\sigma'}\left(\mathbf{r}, t\right),
\end{equation}
where $t$ is time, $\phi_{m\sigma}$ is the component of the Kohn-Sham wavefunction of one-particle state $m$ corresponding to spin $\sigma$ and $H_{KS}$ is the Kohn-Sham Hamiltonian including the kinetic energy, external potential $v_{\mathrm{ext}}$ describing electron interactions with ions and external fields and Hartree, $v_{\mathrm{H}}$, and exchange-correlation, $v_\mathrm{xc}$, terms describing interactions between electrons:
\begin{equation} \label{eq2}
\begin{split}
H_{KS,\sigma\sigma'}=&-\frac{\hbar^2}{2m_e}\nabla^2\delta_{\sigma\sigma'}+v_{\mathrm{ext},\sigma\sigma'}(\mathbf{r}, t)\\&+v_{\mathrm{H}}(\mathbf{r}, t)\delta_{\sigma\sigma'}+v_{\mathrm{xc},\sigma\sigma'}(\mathbf{r}, t).
\end{split}
\end{equation}
Here $m_e$ is the electron mass and $\delta_{\sigma\sigma'}$ is the Kronecker delta. 

The Kohn-Sham wavefunctions provide no information on the real wavefunctions of the system. Nevertheless, they allow to reconstruct the real electron density. In the general non-collinear case, the spin density is introduced as 
\begin{equation} \label{eq3}
n_{\sigma\sigma'}(\mathbf{r}, t)=\sum_m^\mathrm{occ} \phi_{m\sigma}^*\left(\mathbf{r}, t\right)\phi_{m\sigma'}\left(\mathbf{r}, t\right).
\end{equation}
and the Hartree and exchange-correlation terms are considered as functionals of this density.

Following this logic, an analogue of the Dexter exciton coupling from Eq. \ref{Dexter} for TDDFT can be proposed in the form 
\begin{equation} \label{Dexter_tddft}
\begin{split}
V_\mathrm{D}=e^2\sum_{\sigma,\sigma',\tau,\tau'} &\int\int \mathrm{d}\mathbf{r}\mathrm{d}\mathbf{r}' \rho^\mathrm{T*}_\mathrm{A,\sigma\sigma'} (\mathbf{r}) \\&\times g_{\mathrm{xc},\sigma\sigma'\tau\tau'}(\mathbf{r},\mathbf{r}') \rho^\mathrm{T}_\mathrm{D,\tau\tau'} (\mathbf{r}') ,
\end{split}
\end{equation}
where $\rho^\mathrm{T}_\mathrm{\sigma\sigma'}(\mathbf{r})=\gamma^\mathrm{T}_\mathrm{\sigma\sigma'}(\mathbf{r},\mathbf{r})$ and
\begin{equation} \label{xc_kernel}
g_{\mathrm{xc},\sigma\sigma'\tau\tau'}(\mathbf{r},\mathbf{r}')= \frac{\delta v_{\mathrm{xc},\sigma\sigma'}(\mathbf {r})}{\delta n_{\tau\tau'}(\mathbf {r}^{\prime})}
\end{equation}
is the exchange-correlation kernel. 

In the present paper, however, we limit consideration of the Dexter mechanism to singlet excitons without account of spin-orbit coupling when   $g_{\mathrm{xc},\sigma\sigma'\tau\tau'} = g_{\mathrm{xc},\sigma\tau}\delta_{\sigma\sigma'}\delta_{\tau\tau'}$ and the corresponding exciton coupling is given by:
\begin{equation} \label{Dexter_tddft1}
V_\mathrm{D}=e^2 \int\int \mathrm{d}\mathbf{r}\mathrm{d}\mathbf{r}' \rho^\mathrm{T*}_\mathrm{A} (\mathbf{r}) g_{\mathrm{xc}}(\mathbf{r},\mathbf{r}') \rho^\mathrm{T}_\mathrm{D} (\mathbf{r}') ,
\end{equation}
where $g_{\mathrm{xc}}=(g_{\mathrm{xc},\uparrow\uparrow}+g_{\mathrm{xc},\uparrow\downarrow})/2$.
This expression was formally proved in paper \cite{Hsu2001}. Note that here we neglect the term corresponding to the overlap between the donor and acceptor transition densities \cite{Curutchet2017, Mennucci2011, Hsu2001}. 

Therefore, in TDDFT, both the Dexter and F\"orster contributions to the exciton coupling are determined by the transition densities $\rho^\mathrm{T}_{\mathrm{A}}$ and $\rho^\mathrm{T}_{\mathrm{D}}$ of the acceptor and donor. For the IDA coupling described by Eq. (\ref{IDA}), it is sufficient to know the transition dipole moments. Since the ground and excited-state wavefunctions are not known in TDDFT, these quantities are not accessible directly from Eqs. (\ref{gamma}) and (\ref{dipole}).
However, as it turns out \cite{Thiele2009,Hofmann2012,  Fischer2015,Schelter2018,Jornet-Somoza2019}, they can be extracted from the linear-response calculations. 

The linear response $\delta n (\mathbf{r}, t)=\sum_\sigma\delta n_{\sigma\sigma} (\mathbf{r}, t)$  of the charge density to an external potential $\delta v_{\mathrm{ext}}(\mathbf{r}', t')=\sum_\sigma\delta v_{\mathrm{ext},\sigma\sigma} (\mathbf{r}, t)$ is determined by the density response function $\chi(\mathbf{r},\mathbf{r}', t-t')$ \cite{Marques2012,Ullrich2012}:
\begin{equation} \label{response}
\delta n (\mathbf{r}, t) = \sum \int \mathrm{d}t'  \int \mathrm{d}\mathbf{r}'  \chi (\mathbf{r},\mathbf{r}', t-t') \delta v_{\mathrm{ext}}(\mathbf{r}', t'),
\end{equation}
which in the frequency domain corresponds to
\begin{equation} \label{response_w}
\begin{split}
\delta n (\mathbf{r}, \omega) =\int \mathrm{d}\mathbf{r}'   \chi (\mathbf{r},\mathbf{r}', \omega) \delta v_{\mathrm{ext}}(\mathbf{r}',  \omega).
\end{split}
\end{equation}

The density response function can be written in the Lehmann representation \cite{Marques2012,Ullrich2012} as
\begin{equation} \label{response_L}
\begin{split}
\chi =\frac{1}{\hbar} \lim_{\eta\to +0} &\sum_{m=1}^{\infty} \Bigg(\frac{\langle 0 | \hat{n}(\mathbf{r}) |m \rangle  \langle m | \hat{n}  (\mathbf{r}')|0\rangle}{\omega-\Omega_m+i\eta}\\&-\frac{\langle 0 | \hat{n} (\mathbf{r}') |m \rangle  \langle m | \hat{n}(\mathbf{r})|0\rangle}{\omega+\Omega_m+i\eta}\Bigg),
\end{split}
\end{equation}
where $|m\rangle$ denotes the $m$-th excited state wavefunction, $|0\rangle$ is the ground state wavefunction, $\Omega_m$ is the frequency of the $m$-th excitation and $\hat{n}(\mathbf{r}) =\sum_N \delta(\mathbf{r}-\mathbf{r}_i)$ is the density operator.

The resonant part of the density response function thus reads 
\begin{equation} \label{response_L1}
\begin{split}
\chi^\mathrm{res} =-\frac{i \pi}{\hbar}  \sum_{m=1}^{\infty}   &f_0(\omega-\Omega_m)\langle 0 | \hat{n}(\mathbf{r}) |m \rangle \langle m | \hat{n}(\mathbf{r}')|0\rangle,
\end{split}
\end{equation}
where $f_0(\omega) = \delta(\omega)$ is the normalized lineshape. Note that when spin-orbit coupling is not taken into account and the matrix elements above are real, this resonant part exactly corresponds to the imaginary part of the response function: $\chi^\mathrm{res}=i \mathrm{Im}\ \chi$ (see Eq. (\ref{response_L})).

Based on Eqs. (\ref{response_w}) and (\ref{response_L1}), the resonant part of the linear response of the charge density $\delta n$ to the small electric field $\delta\mathbf{E}$ at frequency $\omega$ can be written as 
\begin{equation} \label{response_mu}
\begin{split}
 \delta n^\mathrm{res} (\mathbf{r}, \omega)  =& \frac{i \pi}{\hbar}   \sum_{\mu=1}^{3} \sum_{m=1}^{\infty}  \delta E_\mu f_0(\omega-\Omega_m)  \times \\&\int \mathrm{d}\mathbf{r}' \langle 0 | \hat{n} (\mathbf{r}) |m \rangle  \langle m | \hat{d}_{\mu}(\mathbf{r}')|0\rangle,
\end{split}
\end{equation}
where $\hat{\mathbf{d}}(\mathbf{r})=-e\mathbf{r}\hat{n}(\mathbf{r})$ is the dipole moment operator. The first-order derivative with respect to the electric field  $\tilde{n}^\mathrm{res}_{\mu}=\partial n^\mathrm{res}/\partial E_\mu$ then takes the form
\begin{equation} \label{density_L}
 \tilde{n}^\mathrm{res}_{\mu} (\mathbf{r}, \omega) = \frac{i \pi}{\hbar} \sum_{m=1}^{\infty}   f_0(\omega-\Omega_m) \rho^\mathrm{T*}_{m}(\mathbf{r}) d_{m,\mu}^\mathrm{T},
\end{equation}
where $\mathbf{d}^\mathrm{T}_{m} = \int \mathrm{d}\mathbf{r}\langle m | \hat{\mathbf{d}}(\mathbf{r})|0\rangle$ is the transition dipole moment for the $m$-th excited state and $\rho^\mathrm{T}_{m}(\mathbf{r})=\langle m | \hat{n}(\mathbf{r}) |0 \rangle$ is the corresponding transition density. The transition density can thus be found from the linear response of the charge density close to the resonance once the transition dipole moment is known
\begin{equation} \label{trans_density_L}
\rho^\mathrm{T*}_{m}  (\mathbf{r})= -\frac{i\hbar}{\pi f_0(0)} \sum_{\mu=1}^{3} \frac{\tilde{n}^\mathrm{res}_{\mu} (\mathbf{r}, \Omega_m)e_{m,\mu}^\mathrm{T}}{|\mathbf{d}_{m}^\mathrm{T}|}.
\end{equation}
Here, as before, $\mathbf{e}_{m} = \mathbf{d}_{m}/|\mathbf{d}_{m}^\mathrm{T}|$. Again, when spin-orbit coupling is not taken into account,   $\tilde{n}_{\mu}^\mathrm{res}=i\mathrm{Im}\ \tilde{n}_{\mu}$.

The polarizability corresponds to the linear response of the dipole moment and its resonant part is given by
\begin{equation} \label{polarizability_L0}
\begin{split}
\alpha^\mathrm{res}_{\nu\mu}(\omega)&=\partial d^\mathrm{res}_\nu/\partial E_\mu=-e\int \mathrm{d}\mathbf{r}  x_\nu \tilde{n}^\mathrm{res}_{\mu} (\mathbf{r}, \omega) \\& =\frac{i \pi}{\hbar} \sum_{m=1}^{\infty}    f_0(\omega-\Omega_m) d_{m,\nu}^\mathrm{T*} d_{m,\mu}^\mathrm{T}.
\end{split}
\end{equation}
Therefore, transition dipole moments can be found from the polarizability tensor. 

Irrespective of whether spin-orbit coupling is taken into account or no, the resonance part of the diagonal elements of the polarizability corresponds to the imaginary part of these elements: $\alpha^\mathrm{res}_{\nu\mu}\delta_{\nu\mu}=i\mathrm{Im}\ \alpha_{\nu\mu}\delta_{\nu\mu}$. The same also holds for the trace of the polarizability: 
\begin{equation} \label{polarizability_L}
\begin{split}
\frac{1}{3}\mathrm{Im}[\mathrm{Tr}\alpha (\omega)]&=\frac{\pi}{3\hbar} \sum_{m=1}^{\infty}   f_0(\omega-\Omega_m) |\mathbf{d}_{m}^\mathrm{T}|^2\\&=\frac{\pi}{2}\frac{e^2}{m_e}\sum_{m=1}^{\infty}   \frac{f_m}{\Omega_m}  f_0(\omega-\Omega_m),
\end{split}
\end{equation}
where $f_m$ is the oscillator strength of the $m$-th excitation. The latter can be used to compute the magnitudes of transition dipole moments, while diagonalization of the polarizability tensor allows to establish the directions of transition dipole moments. As soon as the transition dipole moments are known, Eq. (\ref{trans_density_L}) can be used to find the transition densities and Eqs. (\ref{FRET}), (\ref{Dexter_tddft1}) and (\ref{IDA}) to compute the exciton coupling. 

\subsection{Transfer rate}
The exciton transfer rate $k$ are computed according to Eq. (\ref{k_F}). Two approaches are used to estimate the spectral overlap $J_\mathrm{DA}$.

The first one is the Marcus rate expression \cite{Marcus1956, Marcus1993, Stehr2016}:
\begin{equation} \label{eq_marcus2}
J_{\mathrm{M}} = \sqrt{\frac{1}{4\pi\lambda k_{\mathrm{B}}T}} \exp{\left( -\frac{\Delta^2}{4\lambda k_{\mathrm{B}}T}\right)}.
\end{equation}
Here $\lambda$ is the reorganization energy, $\Delta = \lambda+ \Delta G=\hbar(\Omega_\mathrm{D}-\Omega_\mathrm{A})$ is the energy shift in the emission peak of the donor and absorption peak of the acceptor with $\Delta G$ being the free energy difference between the final and initial states of the system, $T$ is temperature and $k_{\mathrm{B}}$ is the Boltzmann constant. The reorganization energy, $\lambda$, can be divided into the contributions associated with geometry relaxation of the ground state, $\lambda_\mathrm{gr}$, and of the excited state, $\lambda_\mathrm{ex}$, so that $\lambda=\lambda_\mathrm{gr}+\lambda_\mathrm{ex}$  \cite{Stehr2016}. These two contributions are related to the structural rearrangements of the donor and acceptor upon the exciton transfer, respectively.

The Marcus rate expression is valid in the classical limit $k_{\mathrm{B}}T\geq\hbar\omega$, when vibrations can be treated classically. However, at room temperature, not all vibrations can be described in this way. To check how these vibrations affect the transfer rates, we consider the Levith-Jortner expression \cite{Levich1959, Levich1966, Jortner1976, Stehr2016}, where the vibrations that cannot be treated classically are taken into account through the effective frequency $\omega_\mathrm{eff}$, Huang-Rhys factor $S_\mathrm{eff}$ and quantum reorganization energy $\lambda_\mathrm{q}=\hbar\omega_\mathrm{eff} S_\mathrm{eff}=\lambda-\lambda_\mathrm{c}$ ($\lambda_\mathrm{c}$ is the classical contribution to the reorganization energy):
\begin{equation} \label{eq_levith}
\begin{split}
&J_{\mathrm{LH}} =\sqrt{\frac{1}{4\pi\lambda_\mathrm{c} k_{\mathrm{B}}T}}\times \\&\sum_p  \frac{S_\mathrm{eff}^p}{p!}\exp{\left(-S_\mathrm{eff}\right)} \exp{\left( -\frac{(\Delta-\lambda_\mathrm{q}+p\hbar\omega_\mathrm{eff})^2}{4\lambda_c k_{\mathrm{B}}T}\right)}.
\end{split}
\end{equation}

\begin{figure}
 \centering
 \includegraphics[width=\columnwidth]{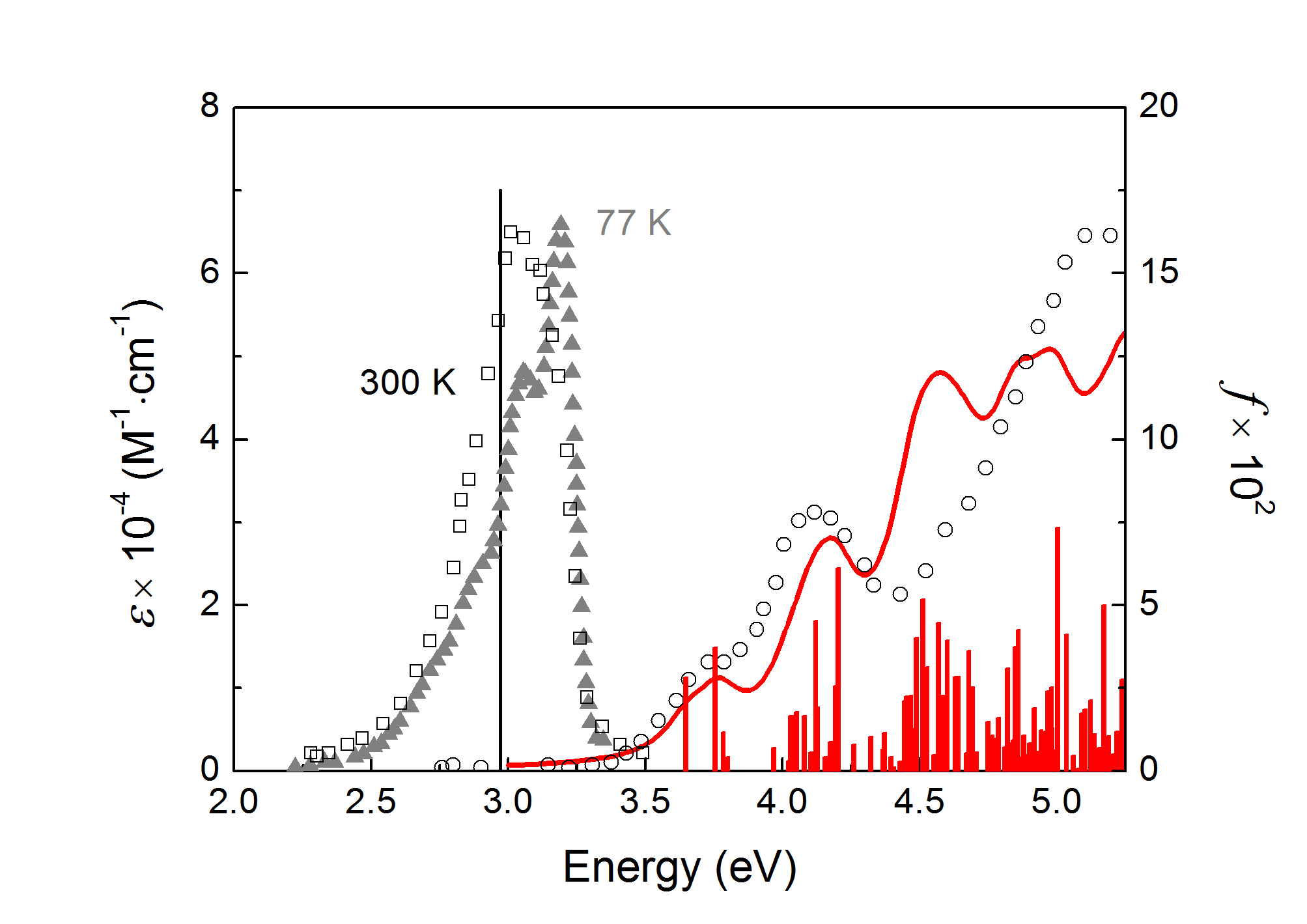}
 \caption{Calculated molar extinction coefficient $\epsilon$ (red solid line, in M$^{-1}\cdot$cm$^{-1}$, left axis) and oscillator strength $f$ of optical transitions (red bars, right axis) for DPBIC as functions of the excitation energy (in eV). The energy corresponding to fluorescence is indicated by the vertical black line. The experimental data \cite{Haneder_tesis} for the absorption spectrum of DPBIC in the polymethylmethacrylate matrix are shown by circles. Experimental photoluminescence spectra at temperatures 77 K \cite{Erk2006} and 300 K \cite{Haneder2008,Haneder_tesis} are shown in arbitrary units by triangles and squares, respectively. The molar extinction coefficient is found from Eq. (\ref{eq6_1}) using the solution of the Casida equation without account of spin-orbit coupling for isolated DPBIC with the PBE0 functional, def2-TZVP basis set and def2/J auxiliary basis set. Lorentzian broadening with half width at half maximum of 0.1 eV is applied. The matrix refractive index is taken equal to 1.5 (polymethylmethacrylate).}
 \label{fig:exp_dpbic} 
\end{figure}

The parameters for this expression are estimated from the vibronic structure of the experimental photoluminescence spectrum of DPBIC at temperature 77 K \cite{Erk2006} (Fig. \ref{fig:exp_dpbic}). The energy corresponding to the effective frequency of the vibrations coupled to the excitation, $\hbar\omega_\mathrm{eff}$, is found to be about 0.138 eV from the energy gap between the first and second peaks in the spectra \cite{Wang2004}. This energy is in the range of values of 0.08 -- 0.14 eV obtained for other iridium complexes \cite{Vries2019}. The values of  $\lambda_\mathrm{q,gr}\approx0.10$ eV and $S_\mathrm{eff,gr} = \lambda_\mathrm{q,gr}/ (\hbar\omega_\mathrm{eff})\approx0.73$ are obtained by fitting the photoluminescence spectrum assuming that the density of states for the donor is described by the equation similar to Eq. (\ref{eq_levith}) (Ref. \onlinecite{Stehr2016}). Using the same quantum contribution to the reorganization energy for the acceptor, $\lambda_\mathrm{q,ex}=\lambda_\mathrm{q,gr}=\lambda_\mathrm{q}/2$, we get total
$\lambda_\mathrm{q}=0.20$ eV and $S_\mathrm{eff}=1.46$. Note that at 300 K, the vibronic structure of the photoluminescence spectrum\cite{Haneder2008,Haneder_tesis} is not resolved and the shape of the spectrum approaches that described by the classical Marcus rate expression (Fig. \ref{fig:exp_dpbic}). Such a qualitative change in the spectrum shape is reproduced using the estimated parameters.

When several optical transitions are involved in the exciton transfer, the characteristic total transfer rate is estimated from the rates of these transitions, $k_i$, as
\begin{equation} \label{eq_marcus3}
k_\mathrm{tot} =\left(\sum_i k_i^2\right)^{1/2}.
\end{equation}

\subsection{Computational details}
\subsubsection{Casida equation} 
To compute absorption spectra and transition dipole moments of isolated species as well as to perform geometry optimization of the first excited states, we use the Casida equation \cite{Casida}. In this approach, poles and residues of the first-order response function are found by solving the eigenvalue problem in the basis of particle-hole states. 

The polarizability is obtained from excitation frequencies and oscillator strengths according to Eq. (\ref{polarizability_L}). The absorption cross-section is related to the polarizability as
\begin{equation} \label{eq6}
\sigma(\omega)=\frac{4\pi\omega}{3c} \mathrm{Im}[\mathrm{Tr}\alpha (\omega)],
\end{equation}
where $c$ is the speed of light, 
and the molar extinction coefficient for molecules in a matrix is computed as
\begin{equation} \label{eq6_1}
\epsilon(\omega)=\frac{4\pi\omega N_\mathrm{A}}{3n_\mathrm{M}c \ln{10}} \mathrm{Im}[\mathrm{Tr}\alpha (\omega)],
\end{equation}
where $N_\mathrm{A}$ is the Avogadro constant and $n_\mathrm{M}$ is the matrix refractive index.

We take the advantage of the efficient implementation of the Casida equation in the ORCA code \cite{Neese2012,Neese2018} based on atom localized basis sets. Balanced polarized triple-zeta basis sets (def2-TZVP) \cite{Weigend2005} and auxiliary basis sets def2/J \cite{Weigend2006} are employed. All electrons are explicitly considered for hydrogen, nitrogen and carbon, while for iridium the quasirelativistic pseudopotential \cite{Andrae1990} is used to replace the core electrons with an effective core potential. The hybrid PBE0 exchange-correlation functional \cite{Adamo1999} is applied. The use of the hybrid functional allows us to improve description of excitation energies and oscillator strengths as compared to the results obtained in the local density approximation (LDA) \cite{Ceperley1980} and generalized gradient approximation (GGA) with the exchange-correlation functional of Perdew, Burke and Ernzerhof (PBE) \cite{Perdew1996} (see Supplementary Material). The resolution of identity (RI) approximation \cite{Eichkorn1995} for evaluation of the Coulomb terms and ``chain of spheres exchange" (COSX) algorithm \cite{Neese2009} for the calculation of the exchange terms (together RIJCOSX \cite{Petrenko2011}) are applied to reduce the computational cost. The single-point energies are converged in self-consistent field calculations within $3\cdot10^{-7}$ eV. The geometry optimization is performed with  the quasi-Newton optimizer using the Broyden-Fletcher-Goldfarb-Shanno (BFGS) update. The forces are converged within $2\cdot10^{-2}$ eV/\AA. The Casida equation \cite{Casida} is solved in the Tamm-Dancoff approximation\cite{Hirata1999}, which is similar in accuracy to full TDDFT for vertical excitations\cite{Hirata1999, Isegawa2013, Grimme2013}. The Davidson diagonalization procedure \cite{Hsu1976} is used. The convergence tolerance for the energies is $10^{-5}$ eV. 

\subsubsection{Real-time propagation} 
The Casida equation provides an efficient tool for calculation of transition dipole moments of isolated species for singlet excitations. This method, however, is based on the use of real wavefunctions and this excludes the possibility of consideration of non-collinear spins. Furthermore, it is not straightforward to decompose the total response into contributions from different spatial domains in this case. To study exciton transfer in supramolecular systems in the present paper, we use the real-time propagation approach as implemented in the Octopus code \cite{Marques2003,Castro2006,Andrade2015,Castro2004a}. The advantage of this method is that the corresponding routines are massively parallelized \cite{Andrade2012} and do not require computation of virtual states. 

In real-time propagation, the time-dependent Kohn-Sham equations (\ref{eq1})--(\ref{eq3}) are solved by explicit integration in time. Initially the system in the ground state is considered. An infinitesimal electric field pulse $\delta E_\mu (t) =\kappa_\mu \delta(t)$, where $\delta(t)$ is the delta function, is applied to simulate the dipole response. The pulses in three orthogonal directions are studied  to extract the full polarizability tensor. The time-dependent dipole moment along the axis $\nu$ is found from the response of the electron density determined by Eq. (\ref{eq3}) as
\begin{equation} \label{eq4}
d_{\nu}(t)=-e\sum_\sigma\int d\mathbf{r} x_\nu n_{\sigma\sigma}(\mathbf{r}, t),
\end{equation}

The Fourier transform of the time-dependent dipole moment gives the polarizability
\begin{equation} \label{eq5}
\alpha_{\nu\mu}(\omega)=\frac{1}{\kappa_\mu}\int_0^{\infty} dt e^{-i\omega t}(d_{\nu}(t)-d_{\nu}(0)).
\end{equation}

To extract the contributions of individual molecules to the total absorption spectrum as well transition densities and transition dipole moments for each of them, the system is decomposed into local domains \cite{Jornet-Somoza2015,Jornet-Somoza2019}. The partitioning is performed on the basis of the quantum theory of atoms in molecules (QTAIM) \cite{Bader1994}. According to this theory, different regions of space are assigned to specific atoms following the gradient of the electron density. The calculations of  time-dependent dipole moments, polarizabilities and cross-sections can then be performed within each local domain according to Eqs. (\ref{eq4})--(\ref{eq6}). 

The individual transition dipole moments computed in this way are used to obtain the IDA exciton coupling  with account of the effects of the local chemical environment \cite{Jornet-Somoza2019} according to Eq. (\ref{IDA}). The individual transition densities are used to evaluate the exciton coupling with account of multipoles beyond the dipole-dipole term according to Eqs. (\ref{FRET}) and (\ref{Dexter_tddft1}). Following the transition density cube method (TDCM) \cite{Krueger1998, Curutchet2017, Jornet-Somoza2019}, integration is performed explicitly on a three-dimensional real-space grid by summing up contributions from finite-size volume elements. 

Gaussian broadening is applied in real-time propagation calculations. The corresponding lineshape is given by
\begin{equation} \label{eq_Gauss}
f_0^\mathrm{G}(\omega) =\frac{1}{\sqrt{2\pi}\Gamma } \mathrm{exp}\left[-\frac{\omega^2}{2\Gamma^2}\right],
\end{equation}
where $\Gamma$ is expressed through the half width at half maximum as $\Gamma= \mathrm{HWHM}/\sqrt{2\mathrm{ln}2}$. The peak height in this case is given by $f_0^\mathrm{G}(0) =1/ (\sqrt{2\pi}\Gamma)$. 

In the Octopus code \cite{Marques2003,Castro2006,Andrade2015,Castro2004a}, one-particle wavefunctions and electron density are discretized on a dense real-space regular grid. The minimum mesh consisting of spheres around each atom is used in the present paper. The grid spacing of 0.14 \AA~and spheres around each atom of the radius of 3.5 \AA~are sufficient to converge the excitation energies within the accuracy of 0.1 eV. The LDA functional \cite{perdew1981} is used. The relativistic separable pseudopotentials of Hartwigsen, Goedecker and Hutter \cite{Hartwigsen1998} are applied to enable calculations with account of spin-orbit coupling. Note that although the excitation energies are strongly underestimated within LDA, the calculated spectra are qualitatively similar to those obtained with the PBE0 functional (see Supplementary Material). Therefore, LDA can be used to qualitatively study exciton dynamics. The 
conjugate-gradients solver\cite{jiang2003} with a mixing parameter of 0.3 is used for the ground-state calculation of eigenstates with the relative tolerance of $10^{-11}$. The approximately enforced time-reversal symmetry (AERTS) propagator \cite{Castro2004a} is used to integrate Kohn-Sham equations (\ref{eq1}) in time. The time step is 0.79 as. The propagation time is almost 200 fs. The strength of the electric field pulse is $\kappa_\mu=\kappa=0.001$ \AA$^{-1}$. This strength is small enough to stay in the linear-response regime. The electronic density is saved every 100 time steps. 

\subsubsection{Sternheimer equation} 
A clear advantage of the real-time propagation approach is that the optical response is obtained for all frequencies at once. The linewidth, however, is determined by the propagation time and long simulation times are required to resolve strongly overlapping bands. Resolving triplet peaks located in the close vicinity of singlet ones with a large oscillator strength can be particularly challenging. 

The calculation of absorption spectra with account of spin-orbital coupling within narrow spectral regions is performed in the present paper based on the Sternheimer equation \cite{Bitzek2006,Andrade2007,Strubbe_tesis,Strubbe2012,Andrade2015,Lebedeva2019} for the first-order derivatives of Kohn-Sham states $\tilde{\phi}_{m\sigma,\mu} = \partial \phi_{m\sigma} /\partial E_\mu$ with respect to the electric field:
\begin{equation} \label{eq_stern}
\begin{split}
\big((\pm\omega +& i\eta-\epsilon_{0,m})\delta_{\sigma\sigma'} + H_{0,\sigma\sigma'}(\mathbf{r})\big) \tilde{\phi }_{m\sigma',\mu} (\mathbf{r},\pm\omega)\\& =-P_c \tilde{H}_{\sigma\sigma',\mu} (\mathbf{r},\pm\omega)  \phi_{0,m\sigma'}(\mathbf{r}). 
\end{split}
\end{equation}
Here $H_{0,\sigma\sigma'}$ is the unperturbed Hamiltonian, $ \epsilon_{0,m}$ is the energy of the unperturbed $m$-th Kohn-Sham state, $P_c$ is the projector onto the unoccupied states, $i\eta$ is a small but finite imaginary frequency added to avoid divergences at the resonances \cite{Andrade2007,Strubbe_tesis,Strubbe2012,Andrade2015} and $\tilde{H}_{\sigma\sigma',\mu}$ is the first-order derivative of the Hamiltonian:
\begin{equation} \label{eq_stern_1}
\begin{split}
\tilde{H}_{\sigma\sigma',\mu}& (\mathbf{r},\omega)=-e x_\mu \delta_{\sigma\sigma'} +e^2\sum_{\tau,\tau'}\int d\mathbf{r'}  \\ &\tilde{n}_{\tau\tau',\mu}(\mathbf{r'},\omega) \left(\frac{\delta_{\sigma\sigma'}\delta_{\tau\tau'}}{|\mathbf{r'}-\mathbf{r}|} +g_{\mathrm{xc},\sigma\sigma'\tau\tau'}(\mathbf{r},\mathbf{r'})\right).
\end{split}
\end{equation}
In our calculations, however, we neglect the exchange-correlation term with $g_{\mathrm{xc}}$ as it normally provides a minor correction to the absorption spectra. The linear-response of the spin density $\tilde{n}_{\sigma\sigma',\mu}$ that enters into the first-order derivative of the Hamiltonian is computed as
\begin{equation} \label{eq_stern_2}
\begin{split}
\tilde{n}_{\sigma\sigma',\mu}(\mathbf{r},\omega)=\sum_m^\mathrm{occ}&\Big[\phi_{0,m\sigma}^{*}(\mathbf{r})\tilde{\phi }_{m\sigma',\mu}(\mathbf{r},\omega) \\&+ \tilde{\phi }^*_{m\sigma,\mu}(\mathbf{r},-\omega)\phi_{0,m\sigma'}(\mathbf{r})\Big]
\end{split}
\end{equation}
and the polarizability tensor is found as
\begin{equation} \label{eq_stern_3}
\alpha_\mathrm{\nu\mu}(\omega)=-e\sum_\sigma \int \mathrm{d}\mathbf{r}  x_\nu \tilde{n}_{\sigma\sigma',\mu} (\mathbf{r}, \omega).
\end{equation}

Since the right-hand side of Eq. (\ref{eq_stern}) depends on derivatives $\tilde{\phi}_{m\sigma,\mu}$ of the Kohn-Sham states through the derivative $\tilde{n}_{\sigma\sigma',\mu}$ of the linear-response charge density given by Eq. (\ref{eq_stern_2}), it has to be solved self-consistently. The resulting polarizability and derivative  $\tilde{n}_{\mu}=\sum_\sigma\tilde{n}_{\sigma\sigma,\mu}$ of the linear-response density allow calculation of the  transition dipole moments according to Eqs. (\ref{polarizability_L0}) and (\ref{polarizability_L}) and transition densities according to Eq. (\ref{density_L}). 

The absorption spectra computed using the Sternheimer approach have Lorentzian broadening with a lineshape
\begin{equation} \label{eq_Lorentz}
f_0^\mathrm{L}(\omega) =\frac{\eta}{\pi} \frac{1}{\omega^2+\eta^2}.
\end{equation}
The value of this function at the peak maximum is $f_0^\mathrm{L}(0)=1/(\pi\eta)$ and $\mathrm{HWHM}=\eta$.

For these calculations, we again use the Octopus code \cite{Andrade2007,Strubbe_tesis,Strubbe2012,Andrade2015,Lebedeva2019,Marques2003,Castro2006,Andrade2015,Castro2004a} and LDA functional. The quasi-minimal residual (QMR) method \cite{Freund1991} (qmr\_dotp) with the final tolerance of $10^{-6}$ is applied to solve linear equations. The procedure similar to the ground-state DFT is applied for self-consistent iterations. The linewidth parameter $\hbar\eta=2$ meV is used.

\section{Electronic structure and absorption spectra}
\subsection{Neutral species in ground-state geometry} 
Let us first consider the electronic structure of the NHC blue emitter and DPBIC. In both of these complexes, the iridium atom has three equivalent ligands (we use notations A, B, C for three fragments of each ligand, see Fig. \ref{fig:fragments}, note that the A and C fragments of the iridium complexes corresponding to the phenylene and phenyl groups, respectively, are the same in the both complexes and B fragments are slightly different). The ideally symmetric structure would have the $C_3$ rotational symmetry and triply degenerate one-particle and molecular states. The DFT optimization of the ground-state geometries, however, shows that the Yann-Teller distortions break the $C_3$ rotational symmetry. Still, as discussed below, there are some one-particle and molecular states that are nearly triply degenerate.

The highest occupied molecular orbitals (HOMO) of the blue emitter and DPBIC look similar (Fig. S2a in Supplementary Material). They have an approximate $C_3$ rotational symmetry and are mostly localized on the metal atom and A fragments of the ligands. The lowest energy unoccupied molecular orbitals (LUMOs) of the iridium complexes, however, are rather different. LUMO, LUMO + 1 and LUMO + 2 orbitals of the NHC blue emitter are nearly degenerate and are mostly localized on the B fragments of the ligands and metal atom. They provide three first optical transitions with very close energies (Fig. \ref{fig:trans_dens}a). In DPBIC (Fig. S2e in Supplementary Material), LUMO has an approximate $C_3$ rotational symmetry and is mostly distributed over the metal atom and C fragments of the ligands. Close to this molecular orbital in energy, there are also three orbitals localized on the B and C fragments and metal atom. These four orbitals are mixed in the first optical transitions (Fig. \ref{fig:trans_dens}e). Despite of these differences, it is seen that both in the emitter and DPBIC, the first optical transitions are characterized by a significant charge transfer between the ligands and from the metal atom to the ligands (Figs. \ref{fig:trans_dens}a and e). 

 \begin{figure*}
 \centering
 \includegraphics[width=0.6\textwidth]{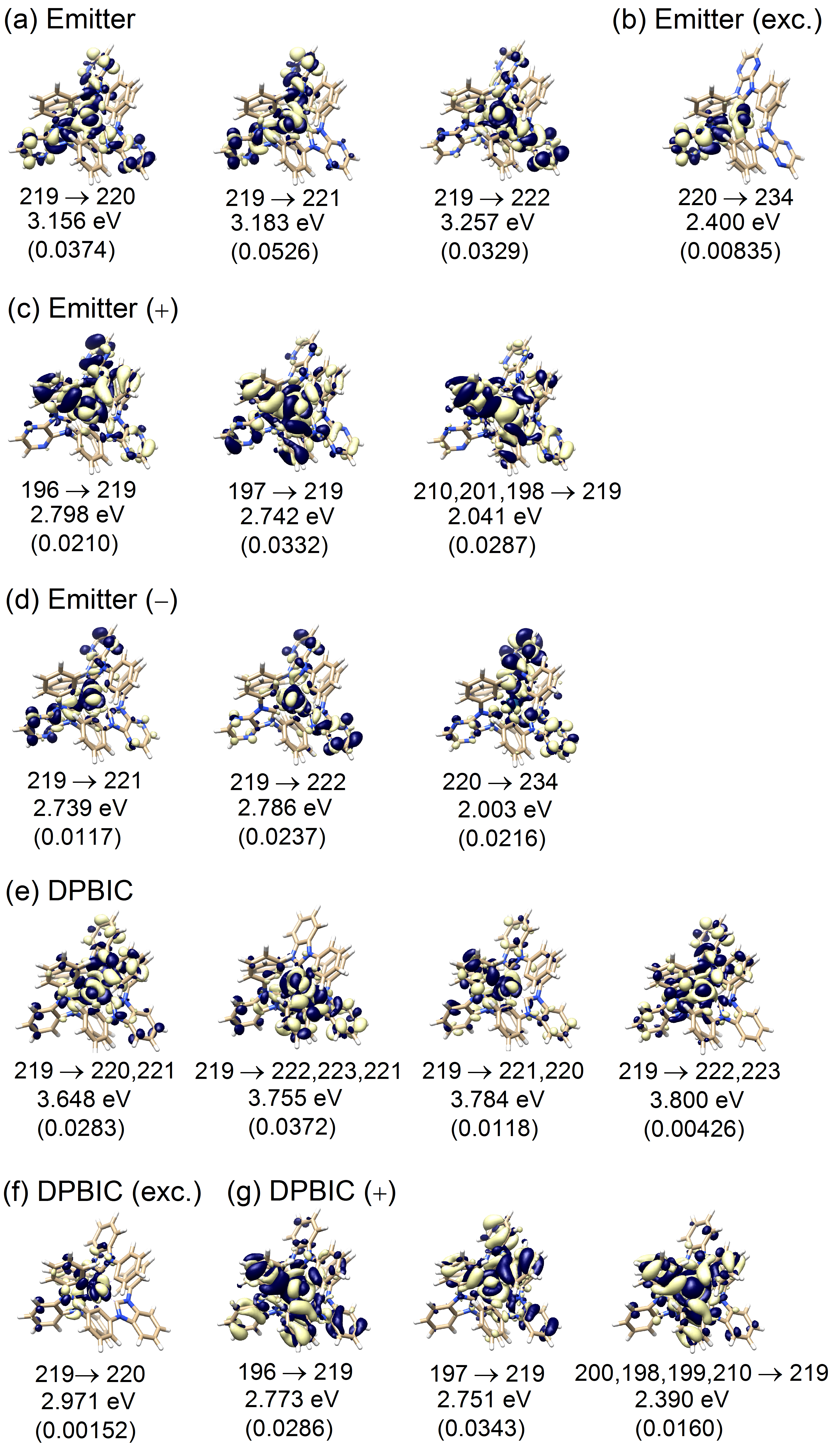}
 \caption{Contour plots of transition densities for (a) ground-state, (b) excited, (c) positively charged and (d) negatively charged NHC blue emitter and (e) ground-state, (f) excited and (g) positively charged DPBIC. For the neutral species, the lowest-energy transitions are shown. For the charged species, bright excitations close to the emission peak of the NHC dye are included. Isosurfaces correspond to 0.0003 \AA$^{-3}$.  The indices of the orbitals taking part in the excitations, excitation energies and oscillator strength (in braces) are shown below the transition densities. Carbon, nitrogen, hydrogen and iridium atoms are coloured in beige, blue, white and sea blue, respectively. The transition densities are obtained by solving the Casida equation without account of spin-orbit coupling for isolated complexes using the PBE0 functional, def2-TZVP basis sets and def2/J auxiliary basis sets.}
 \label{fig:trans_dens} 
\end{figure*}

\begin{figure}
 \centering
 \includegraphics[width=\columnwidth]{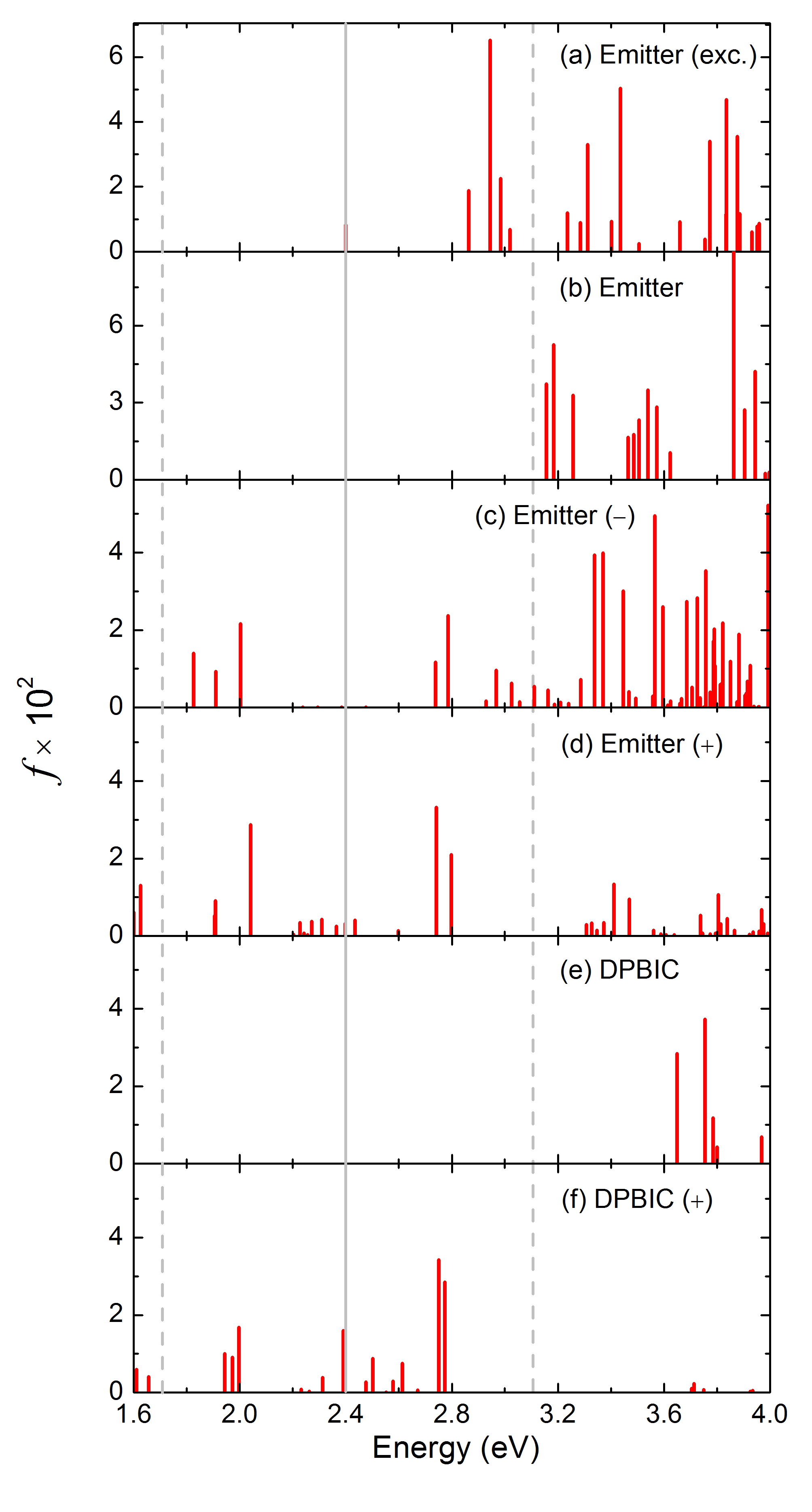}
 \caption{Computed oscillator strengths $f$ of optical transitions at different energies (in eV) for the following species: (a) NHC blue emitter in the excited-state geometry, (b) NHC blue emitter in the ground-state geometry, (c) negatively charged NHC blue emitter, (d) positively charged NHC blue emitter, (e) DPBIC in the ground-state geometry and (f) positively charged DPBIC. The energy corresponding to the emission of the NHC blue emitter of 2.4 eV is indicated by the vertical grey line. The energy interval $2.4\pm0.7$ eV, where the optical transitions in the charged species significantly contribute to the exciton transfer, is shown by the vertical dashed lines. The spectra are obtained by solving the Casida equation without account of spin-orbit coupling for isolated complexes using the PBE0 functional, def2-TZVP basis sets and def2/J auxiliary basis sets.}
 \label{fig:singlet} 
\end{figure}

The absorption spectra for the emitter and DPBIC computed without account of spin-orbit coupling are shown in Figs. \ref{fig:singlet}b and e, respectively. It should be noted that when spectral broadening is taken into account, the calculated absorption spectrum for DPBIC shows a good agreement with the experimental data \cite{Haneder_tesis} (Fig. \ref{fig:exp_dpbic}).

\subsection{Neutral species in excited-state geometry}
Geometry optimization of the first excited states of the NHC blue emitter and DPBIC leads to the decrease of the excitation energy by 0.76 eV and 0.68 eV, respectively. Note that these reorganization energies are larger than typical values found for iridium complexes
of 0.2 -- 0.5 eV \cite{Vries2019}. The two contributions to the reorganization energy, $\lambda_{ex}$ and $\lambda_{gr}$,  related to the structural rearrangements of the acceptor and donor are listed in Table \ref{table:reorg}. 

The calculated energy at which fluorescence takes place for DPBIC of about 3.0 eV agrees well with the positions of the maximum at 3.0 eV \cite{Haneder2008} and 3.2 eV \cite{Erk2006} in the experimentally measured photoluminescence spectra (Fig. \ref{fig:exp_dpbic}) and the optical band gap of 3.2 eV determined experimentally in Ref. \onlinecite{Haneder2008}. Note that the singlet-triplet splitting in this material is small because of the strong spin-orbit coupling. This was shown in \textit{ab initio} calculations \cite{Haneder_tesis, Haneder2008} and is consistent with the experimentally measured small radiative lifetimes \cite{Haneder_tesis}. Therefore, fluorescence and phosphorescence frequencies of DPBIC should be almost the same. The calculated  energy at which fluorescence takes place for the NHC blue emitter of 2.4 eV is also in agreement with the position of the peak maximum in the corresponding electroluminescence spectrum at 2.6 eV\cite{Niu2017,Heimel2018}.

\begin{table}
    \caption{Reorganization energies, $\lambda$,  and contributions to the reorganization energy from the geometry relaxation of the excited state, $\lambda_\mathrm{ex}$, and ground state, $\lambda_\mathrm{gr}$, computed\footnote{The calculations are performed by solving the Casida equation for isolated complexes using the PBE0 functional, def2-TZVP basis sets and def2/J auxiliary basis sets.} for the NHC blue emitter and DPBIC.}
   \renewcommand{\arraystretch}{1.2}
    \begin{ruledtabular}
        \begin{tabular}{*{4}{c}}
Complex & $\lambda$ (eV) & $\lambda_\mathrm{ex}$ (eV) & $\lambda_\mathrm{gr}$ (eV) \\\hline
Blue emitter &  0.756   & 0.3789 & 0.3776\\
DPBIC & 0.677 & 0.2263 & 0.4506 \\
\end{tabular}
\end{ruledtabular}
\label{table:reorg}
\end{table}

 \begin{figure}
 \centering
 \includegraphics[width=0.45\textwidth]{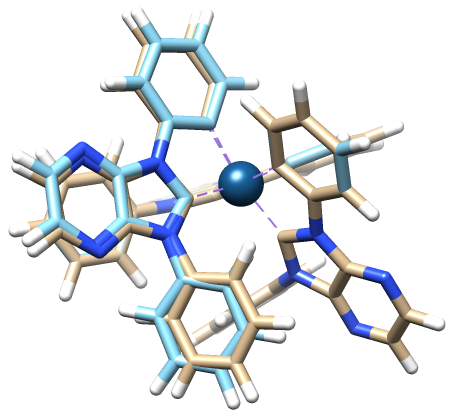}
 \caption{Atomic structures of the optimized ground and excited-state geometries of the NHC blue emitter. Carbon atoms in the ground and excited-state geometries are shown in beige and cyan, respectively. Nitrogen, hydrogen and iridium atoms are coloured in blue, white and sea blue, respectively. The positions of two ligands and the metal atom are the same in the ground and excited-state geometries. The Casida equation is solved without account of spin-orbit coupling for the isolated complexes using the PBE0 functional, def2-TZVP basis sets and def2/J auxiliary basis sets.}
 \label{fig:emitter_exc_geom} 
\end{figure}

In the excited-state geometry, LUMOs of both of the organometallic complexes considered are no longer degenerate and HOMOs are not symmetric (Figs. S2b and f in Supplementary Material). HOMOs are mostly localized on the A fragment of one of the ligands and the metal atom. In the NHC blue emitter, LUMO is centered on the B fragment of the same ligand, while LUMO of DPBIC occupies the C fragment of another ligand that is close to the A fragment carrying HOMO.  The first optical transition of both of the complexes in the excited state geometry, therefore, has a metal-ligand charge transfer (MLCT) character (Figs. \ref{fig:trans_dens}b and f). For DPBIC, where the MLCT character is particularly pronounced, this conclusion is supported by the poorly resolved vibronic structure of the emission spectrum \cite{Haneder_tesis, Haneder2008} (see also Fig. \ref{fig:exp_dpbic}). A consequence of the strong MLCT character of the first optical transition is that reorganization of the complex geometry upon the excitation is accompanied by a significant decrease in the transition dipole moment and oscillator strength. For the NHC blue emitter, there is a two-fold reduction in the transition dipole moment and more than a four-fold drop in the oscillator strength (Figs. \ref{fig:trans_dens}a, d and \ref{fig:singlet}a, b).  For DPBIC, the changes in the transition dipole moment and oscillator strength of the first optical transition are even more dramatic. The transition dipole moment decreases by a factor of 4 and the oscillator strength almost by a factor of 20 (Figs. \ref{fig:trans_dens}a and c).

It should be mentioned that for the NHC blue emitter, the geometry relaxation induced by the excitation is mostly related to structural reconstruction of only one ligand (Fig. \ref{fig:emitter_exc_geom}). When two other ligands, the metal atom and dihedral angles within the optimized ligand are fixed during the geometry optimization of the excited state starting from the ground-state geometry, the reorganization energy of the emitter complex decreases only by 0.07 eV. This can be attributed to the localization of HOMO and LUMO on the same ligand of the NHC blue emitter. For DPBIC, two ligands are involved in the geometry relaxation upon the excitation in accordance with the distribution of HOMO and LUMO (Fig. S2f in Supplementary Material).

As seen from the absorption spectra shown in Figs. \ref{fig:singlet}a and e, there is a large energy gap (1.25 eV) between the emission of the NHC blue emitter, which happens at the energy of the first absorption peak in the excited-state geometry, and absorption edge of DPBIC. Therefore, the transfer of excitons from the emitter complex to DPBIC is unlikely. The significant difference in the positions of the absorption edge and emission peak of the NHC blue emitter (Figs. \ref{fig:singlet}a and b) can also lead to suppression of the exciton transport between emitter complexes.

\subsection{Charged species}
Let us now discuss whether the exciton transfer is possible to charged complexes within the emission layer, i.e. to negatively and positively charged NHC blue emitter and positively charged DPBIC. The absorption spectra of these species in the ground-state geometry are shown in Figs. \ref{fig:singlet}c, d and f and all of them have an overlap with the first absorption peak of the neutral NHC emitter in the excited-state geometry responsible for emission. 

We assume that the reorganization energy for the charged complexes is the same as for the neutral ones (in fact our calculations of the transfer rates reveal that they are insensitive to the reorganization energies used for charged species). Therefore, the most relevant optical transitions of the charged species we should analyze lie in the interval of excitation energies of $2.4 \pm 0.7$ eV centered at the emission peak of the neutral NHC blue emitter (Figs. \ref{fig:singlet}, \ref{fig:trans_dens}b,c). 

In the negatively charged blue emitter, one of the LUMOs of the emitter gets occupied. HOMO, LUMO+1 and LUMO+2, however, still resemble those in the neutral complex and the transitions between them result to be within the energy interval of interest (peaks at 2.73 eV and 2.79 eV in Fig. \ref{fig:trans_dens}d). The contribution of these optical transitions to the exciton transfer is expectable. At low excitation energies within the considered energy interval (Fig. \ref{fig:singlet}c), there are also optical transitions that are not present in the neutral complex: from LUMO to high-energy states of the neutral molecule  (Fig. \ref{fig:trans_dens}d). The oscillator strengths of these transitions are comparable to the transitions HOMO$\to$LUMO+1 and HOMO$\to$LUMO+2. 

In the positively charged blue emitter, the optical transitions within the energy interval of interest are related to transitions from low-lying states to the lowest unoccupied state coming from HOMO of the neutral complex (Fig. \ref{fig:trans_dens}c). The oscillator strengths of some of these transitions (like between molecular orbitals 196$\to$219 and 197$\to$219) are on the order of those for the relevant optical transitions in the negatively charged blue emitter. However, in addition, there are plenty of optical transitions with smaller oscillator strengths close in energy to the emission peak of the NHC blue emitter (Fig. \ref{fig:singlet}d). In spite of the reduced oscillator strengths, it can be expected that such transitions contribute remarkably to the exciton transfer because of the smaller energy shift. 

Similar observations can be also done for optical transitions in positively charged DPBIC (Figs. \ref{fig:trans_dens}g and \ref{fig:singlet}f). A bright example is that similarities in molecular orbitals 196, 197 and 219 (HOMO of neutral species) of the NHC blue emitter and DPBIC provide very close frequencies, oscillator strengths and transition densities for optical transitions 196$\to$219 and 197$\to$219 in these two complexes (Figs. \ref{fig:trans_dens}d and g). The same as in the positively charged blue emitter, in positively charged DPBIC, there are also a number of optical transitions with comparable or smaller oscillator strengths close in energy to the emission peak of the NHC blue emitter (Fig. \ref{fig:singlet}f).

\section{Triplet spectra}

\begin{figure}
 \centering
 \includegraphics[width=\columnwidth]{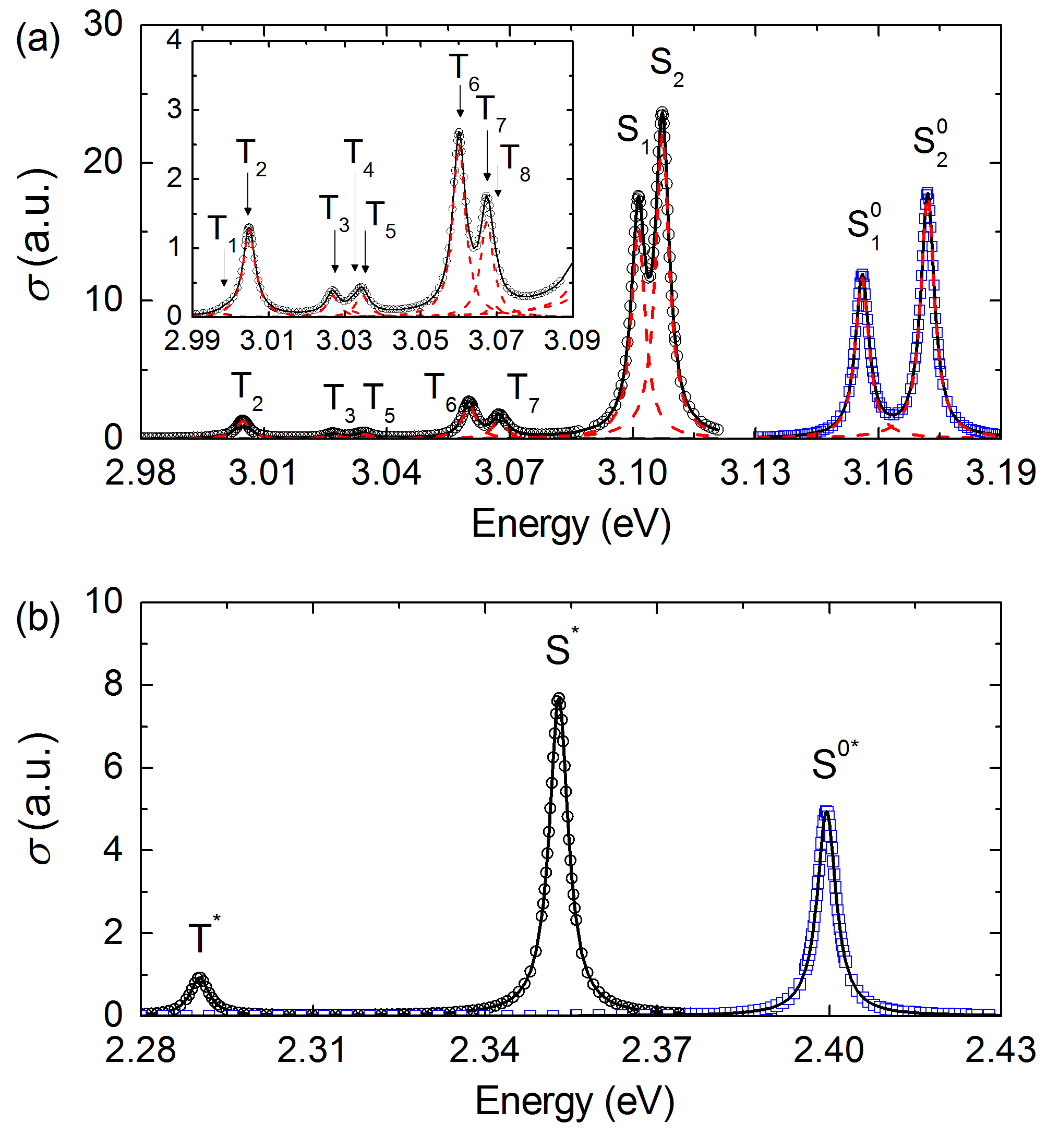}
 \caption{Absorption cross-sections (in a. u.) of the NHC blue emitter in the ground-state (a) and excited-state (b) geometries computed with (black circles) and without (blue squares) account of spin-orbit coupling. The insert in panel (a) shows the part of the spectrum corresponding to triplet excitons with a higher resolution. Approximations by sums of Lorentzian functions are shown by black solid lines. Contributions of individual Lorentzian functions are indicated by red dashed lines. The peaks corresponding to singlet and triplet excitons are denoted by letters S and T, respectively, and are numbered consecutively. The peaks for the excited-state geometry are marked by an asterisk. The peaks obtained without account of spin-orbit coupling are denoted as S$^0$. The Sternheimer equation with the linewidth parameter of $\hbar\eta=2$ meV is solved for the isolated complexes using the regular real-space grid with spacing of 0.14 \AA~and the LDA functional. The scissor operator is applied to correct the band gap. }
 \label{fig:triplet} 
\end{figure}

The absorption spectra of the NHC blue emitter in the ground and excited-state geometries obtained with account of spin-orbit coupling are shown in Figs. \ref{fig:triplet}a and b, respectively. These spectra have been computed using the Sternheimer approach and LDA functional (see section II.D3). For comparison, we also show the spectra obtained with the same approach but without account of spin-orbit coupling (the corresponding peaks are marked here and below as $S_m^0$, where $m$ is the peak index). Since the LDA functional is known to strongly underestimate the band gap, we apply the scissor operator, i.e. rigidly shift the spectra to higher energies changing the cross-section proportional to energy (that is keeping the values of transition dipole moments, see Eqs.  (\ref{polarizability_L}) and (\ref{eq6})). The blue shift of 0.95 eV is used for the ground-state geometry and 0.86 eV for the excited-state one to reproduce positions of the first peaks from the PBE0 calculations without account of spin-orbit coupling (Figs. \ref{fig:singlet}a,b and Fig. \ref{fig:triplet}).

From Fig. \ref{fig:triplet}, it follows that the energy difference between the first peaks corresponding to singlet and triplet excitons is about 0.10 eV and 0.06 eV for the ground and excited-state geometries, respectively. Therefore, our results confirm that the the singlet-triplet splitting is very small for the NHC blue emitter. The amplitudes of the highest triplet peaks (T$_2$, T$_6$ and T$_7$ for the ground-state geometry and T$^*$ for the excited state one) are an order of magnitude smaller than those of the first singlet peaks. Thus, it can be expected that transport phenomena involving triplet excitons occur slower compared to those for singlet excitons. 

From comparison of the data obtained with and without account of spin-orbit coupling in Fig. \ref{fig:triplet}, it is seen that account of spin-orbit coupling leads to changes in the peaks corresponding to singlet excitons. A small red shift of these peaks by 0.05--0.06 eV is observed. Additionally, their amplitudes are increased by 30\% and 60\% for the ground and excited-state geometries, respectively. 

To extract the transition dipole moments, the spectra are fitted by a sum of Lorentzian functions given by Eq. (\ref{eq_Lorentz}):
\begin{equation} \label{eq_Lsum}
\sigma(\omega)=\sum_{m=1}^{M}\sigma_m f_0^\mathrm{L}(\omega-\Omega_m).
\end{equation}
As seen from Eqs.  (\ref{polarizability_L}) and (\ref{eq6}),  parameters $\sigma_m$ can be expressed through the oscillator strength or transition dipole moment as $\sigma_m=2\pi^2e^2f_m/(m_ec)=4\pi^2\Omega_m|\mathbf{d}_{m}^\mathrm{T}|^2/(3\hbar c)$. The parameters of Eq. \ref{eq_Lsum} that minimize the root-mean-square deviation from the computed data as well as estimated transition dipole moments are given in Table \ref{table:triplet_sp}.

The transition dipole moment for the emitter in the excited state can be used to estimate the exciton radiative lifetime according to the equation \cite{Mori2014, Vries2019, Vries2020, Strickler2004, Nozaki2006}
\begin{equation} \label{eq_RLT}
\frac{1}{\tau_\mathrm{r,m}} = \frac{4\epsilon_\mathrm{r}\Omega_m^3|\mathbf{d}_{m}^\mathrm{T}|^2}{3\hbar c^3}.
\end{equation}
Here $\epsilon_\mathrm{r}$ is the optical dielectric permittivity of the surrounding medium, which typically can be taken equal to $\epsilon_\mathrm{r} = 3$ for OLED emission layers\cite{Salehi2018, Liu2005}.
For triplet excitons (peak T$^*$), this gives $\tau_\mathrm{r,T} \sim 1$ $\mu$s, close to the experimental result for the considered blue emitter of 1.6 $\mu$s \cite{Heimel2018} and within the range of values 0.5 -- 10 $\mu$s obtained for other iridium complexes \cite{Haneder_tesis, Haneder2008, Vries2019, Vries2020, Ligthart2021, Ligthart2018}. For singlet excitons (peak S$^*$), the radiative lifetime is estimated to be $\tau_\mathrm{r,S} \sim 0.1$ $\mu$s. This is not that small as for fluorescent dyes \cite{Giebink2008} because of the relatively low oscillator strength (Table \ref{table:triplet_sp}).

\begin{table}
    \caption{Characteristics of the first optical transitions of the NHC blue emitter in the ground and excited-state geometries obtained by fitting the spectra computed\footnote{The LDA functional, the scissor operator and real-space grid are used.} by solving the Sternheimer equation for isolated complexes with and without account of spin-orbit coupling.}
 \centering
   \renewcommand{\arraystretch}{1.2}
    \begin{ruledtabular}
        \begin{tabular}{*{5}{c}}
Peak & $\hbar\Omega$  (eV) & $\sigma \times 10^3$  (a.u.) & $f \times 10^2$ & $|\mathbf{d}^\mathrm{T}|^2$ (a.u.$^2$) \\\hline

\multicolumn{5}{c}{Ground-state geometry with spin-orbit coupling} \\
T$_1$ & 2.999 & 0.012 & 0.008 & 0.001 \\
T$_2$ & 3.005 & 0.295 & 0.204 & 0.028\\
T$_3$ & 3.027 & 0.073 & 0.050 & 0.007 \\
T$_4$ & 3.032 & 0.020 & 0.014 & 0.002 \\
T$_5$ & 3.035 & 0.076 & 0.053 & 0.007 \\
T$_6$ & 3.060 & 0.578 & 0.401 & 0.054 \\
T$_7$ & 3.067 & 0.321 & 0.223 & 0.030 \\
T$_8$ &  3.069 & 0.040 & 0.028 & 0.004 \\
S$_1$ & 3.101 & 3.524 & 2.446 & 0.322 \\
S$_2$ & 3.107 & 5.098 & 3.539 & 0.465 \\
\multicolumn{5}{c}{Excited-state geometry with spin-orbit coupling} \\
T$^*$ & 2.290 & 0.214 & 0.149 & 0.027 \\
S$^*$ &  2.353 & 1.775 & 1.232 & 0.214 \\
\multicolumn{5}{c}{Ground-state geometry without spin-orbit coupling} \\
S$_1^0$ & 3.156 & 2.694 & 1.870 & 0.242 \\
S$_2^0$ & 3.172 & 4.074 & 2.827 & 0.364 \\
\multicolumn{5}{c}{Excited-state geometry without spin-orbit coupling} \\
S$^{0*}$ & 2.400 & 1.143 & 0.793 & 0.135 \\
\end{tabular}
\end{ruledtabular}
\label{table:triplet_sp}
\end{table}

Let us now discuss the accuracy of the LDA calculations. As seen from Table \ref{table:triplet_sp}, the transition dipole moments for singlet excitons computed using the LDA functional without account of spin-orbit coupling are fairly close to the results of PBE0 calculations. The squares of the transition dipole moments are different by a factor of 2 for the ground-state geometry and by less than 5\% for the excited-state one. It should be also noted that the difference in the positions of the absorption edge and emission peak computed with the LDA functional without account of spin-orbit coupling is less than the same quantity for PBE0 only by 0.1 eV. A qualitative similarity in the results of LDA and PBE0 calculations using the same ORCA code can be also appreciated in Fig. S1 of Supplementary Material. Therefore, the LDA functional can be used for qualitative studies of exciton dynamics.

\section{Dexter vs. F\"orster mechanisms}
\begin{figure}
 \centering
 \includegraphics[width=\columnwidth]{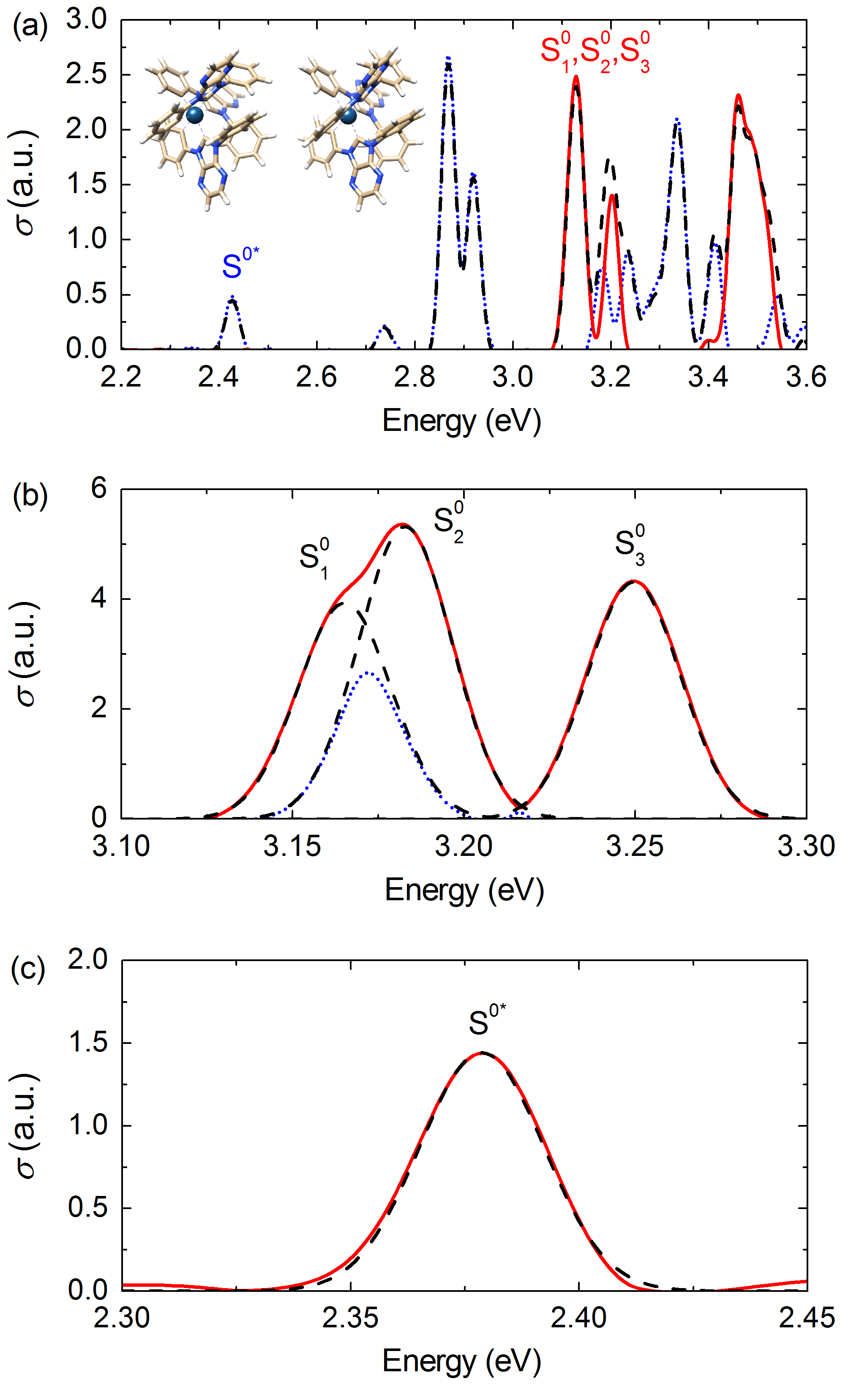}
 \caption{(a) Absorption cross-section (in a. u.) of the dimer consisting of the NHC blue emitters in the ground (acceptor) and excited (donor) states. The acceptor (red solid line), donor (blue dotted line) and total (black dashed line) contributions are shown. The insert demonstrates the atomic structure of the dimer. (b) The largest (red solid line) and the medium (blue dotted line) eigenvalues of the cross-section tensor for the acceptor at the absorption edge. (c) The largest eigenvalue of the cross-section tensor for the donor at the absorption edge. Fitting by individual Gaussian functions is shown in panels (b) and (c) by black dashed lines. The results are obtained by the real-time propagation for the dimer using the regular real-space grid with spacing of 0.14 \AA~and the LDA functional. The scissor operator is applied to correct the band gap. The spin-orbit coupling is not taken into account.}
 \label{fig:dimer} 
\end{figure}

To analyze the contributions of Dexter and F\"orster mechanisms to the exciton coupling, we have performed real-time propagation (see section II.D2) for a dimer consisting of NHC blue emitters in the ground and excited state geometries. The organometallic complexes are separated by 1.2 nm (the distance between the metal atoms), which is a typical distance between them in the considered emission layer\cite{Paterson2019}. The total absorption cross-section for the dimer obtained by real-time propagation is shown in Fig. \ref{fig:dimer}a. Different regions of space are assigned to the donor (emitter in the excited state) or acceptor (emitter in the ground state) based on the QTAIM analysis of the ground-state electron density \cite{Bader1994}. Using such a space division, the contributions of the donor and acceptor to the response electron density and finally to the absorption spectrum (Fig. \ref{fig:dimer}a) are found. It is seen from Fig. \ref{fig:dimer}a that positions of the emission peak S$^{0*}$ and first absorption peaks of the donor, S$_1^0$ and S$_2^0$, are rather close to those obtained using the Sternheimer equation for the isolated emitters in the excited and ground states, respectively (Fig. \ref{fig:triplet}b, Table \ref{table:triplet_sp}).

To resolve the overlapping peaks, we diagonalize the cross-section tensors for the donor and acceptor at each frequency and analyze the cross-section eigenvalues. The frequency-dependent cross-section eigenvalues give the information about the probability
of orthogonal excitations (i.e. transitions characterized by orthogonal transition dipoles), and enable to distinguish between quasi-degenerate states\cite{Jornet-Somoza2019}.
The absorption peaks are clearly seen in the plots of the largest cross-section eigenvalue (Figs. \ref{fig:dimer}b and c) and we approximate them by the sum of Gaussian functions (see Eq. (\ref{eq_Gauss})). For the considered propagation time and damping applied to the spectra, $\mathrm{HWHM}$ is 0.016 eV. The fitted parameters and transition dipole moments estimated according to Eq. (\ref{polarizability_L0}) are listed in Table \ref{table:dimer}. As seen from comparison with Table \ref{table:triplet_sp}, the transition dipole moments obtained by the real-time propagation are about 10\% less than the values from the Sternheimer equation. This decrease can be attributed to the effect of the interaction between the donor and acceptor complexes taken into account explicitly in the real-time propagation \cite{Jornet-Somoza2019}.  

\begin{table*}
    \caption{Characteristics of the first optical transitions of the NHC blue emitter in the ground and excited-state geometries obtained\footnote{The LDA functional, the scissor operator and real-space grid are used.} by fitting the results of the real-time propagation for the emitter dimer.}
 \centering
   \renewcommand{\arraystretch}{1.2}
    \begin{ruledtabular}
        \begin{tabular}{*{8}{c}}
Peak & $\hbar\Omega$  (eV) &  $\sigma \times 10^3$  (a.u.) & $f \times 10^2$ & $|\mathbf{d}^\mathrm{T}|^2$ (a.u.$^2$) & $e_x$  & $e_y$  & $e_z$ \\\hline
\multicolumn{8}{c}{Ground-state geometry} \\
S$_1^0$ & 3.165 &  2.346 & 1.628 & 0.210 & -0.267 & 0.314 & 0.911 \\
S$_2^0$ & 3.183 &  3.168 & 2.199 & 0.282 & -0.170 & -0.554 & 0.815 \\
S$_3^0$ & 3.250 &  2.546 & 1.177 & 0.222 & -0.449 & -0.642 & -0.621 \\
\multicolumn{8}{c}{Excited-state geometry} \\
S$^{0*}$ & 2.380 &  0.941 & 0.653 & 0.112 & 0.144 & -0.9847 & 0.065 \\
\end{tabular}
\end{ruledtabular}
\label{table:dimer}
\end{table*}

After evaluation of the transition dipole moments, the fast Fourier transform of the response electron density is performed at frequencies corresponding to peaks S$^{0*}$, S$_1^0$, S$_2^0$ and S$_3^0$. Based on the assignment of regions of space to the donor and acceptor performed previously by the analysis of the ground-state electron density, the Fourier-transformed densities are also divided into donor and acceptor contributions. Using the transition dipole moments from  Table \ref{table:dimer} (including their directions $\mathbf{e}$), we find the transition densities for the donor at peak S$^{0*}$ and for the acceptor at peaks S$_1^0$, S$_2^0$ and S$_3^0$ according to Eq. (\ref{trans_density_L}). The F\"orster and Dexter exciton couplings computed using the explicit integration over the transition densities (TDCM approach) as well as the IDA F\"orster coupling evaluated according to Eqs. (\ref{FRET}), (\ref{Dexter_tddft1}) and (\ref{IDA}), respectively, are listed in Table \ref{table:dimer_coupling}. 

The Dexter coupling strongly depends on the exchange-correlation kernel $ g_{\mathrm{xc}}$ used for Eq. (\ref{Dexter_tddft1}) and we have computed it for different kernels (Table \ref{table:dimer_coupling}). The Dexter coupling is completely zero for LDA \cite{perdew1981} since in this approximation, the exchange-correlation effects are completely local. The generalized gradient approximation like PBE \cite{Perdew1996} gives the Dexter coupling 3 -- 5 orders of magnitude smaller than the F\"orster one. The PBE0 \cite{Adamo1999} and HSE06 \cite{Heyd2003, Krukau2006} functionals  taking into account a fraction of the exact exchange and non-local effects provide results similar to PBE. Another hybrid functional B3LYP \cite{Lu2013} predicts an increase in the Dexter coupling by an order of magnitude compared to PBE. 
In general, however, it can be concluded that for distances 1.2 nm and more, typical for the OLED emission layer, the Dexter coupling can be neglected in comparison with the F\"orster coupling. Note that kinetic Monte-Carlo simulations \cite{Mesta2016} support that roll-off of white OLEDs occurs via long-range F\"orster mechanism rather than by short-range nearest-neighbour quenching. In the following, we limit our consideration to the F\"orster mechanism.

It is also seen from Table \ref{table:dimer_coupling} that the F\"orster couplings computed taking into account spatial distributions of the transition densities given by Eq. (\ref{FRET}) lie in the same range of magnitudes as the IDA results. Nevertheless, the values for the same peak differ considerably. This is in agreement with results of our previous paper \cite{Jornet-Somoza2019} that IDA is adequate for distances well above 1.2 nm. This result means that accurate simulations of exciton transport in the emission layer require explicit calculations of exciton couplings for diverse acceptor-donor arrangements present in the emission layer, while IDA can only be used for qualitative analysis.

\begin{table*}
    \caption{F\"orster ($V_\mathrm{F}$) and Dexter ($V_\mathrm{D}$) coupling and coupling in the ideal dipole approximation ($V_\mathrm{F}^\mathrm{IDA}$) computed using different exchange-correlation functionals for exciton transfer from the lowest excited state of the excited emitter ($\hbar\Omega_{\mathrm{D}}=2.380$ eV) to the lowest excited states of the emitter in the ground state from the results of real-time propagation\footnote{The LDA functional, the scissor operator and real-space grid are used, spin-orbit coupling is not taken into account.} for the emitter dimer.}
 \centering
   \renewcommand{\arraystretch}{1.2}
    \begin{ruledtabular}
        \begin{tabular}{*{9}{c}}
Peak & $\hbar\Omega$ & $V_\mathrm{F}$ &  \multicolumn{5}{c}{$V_\mathrm{D}$ (meV)} & $V_\mathrm{F}^\mathrm{IDA}$ \\
 & (meV) & (meV) & LDA & PBE & PBE0 & HSE06 & B3LYP  & (meV) \\\hline
S$^0_1$ & 3.165 &  0.494 & 0 & $1.03\cdot10^{-4}$ & $8.31\cdot10^{-5}$ &
$1.29\cdot10^{-4}$ & $1.40\cdot10^{-3}$ & 0.061 \\
S$^0_2$ & 3.183 &  0.570 & 0 & $4.62\cdot10^{-5}$ & $4.00\cdot10^{-5}$ & $1.78\cdot10^{-5}$ & $8.57\cdot10^{-4}$ & 0.269 \\
S$^0_3$ & 3.250 &  0.042 & 0 & $7.73\cdot10^{-5}$ & $4.84\cdot10^{-5}$ & $7.41\cdot10^{-5}$ & $2.22\cdot10^{-2}$ & 0.268 \\
\end{tabular}
\end{ruledtabular}
\label{table:dimer_coupling}
\end{table*}

\section{Transfer rates}
\subsection{Singlet excitons}
Although the IDA (Eq. (\ref{IDA})) fails to describe the exciton coupling between nearest neighbours, it still gives the correct order of magnitude and becomes quantitatively correct at large distances. Therefore, here we use the IDA for qualitative analysis of transfer rates to different species in the emission layer. To take into account the effect of the surrounding dielectric medium, we multiply the exciton coupling in vacuum (Eq. \ref{IDA}) by the screening factor $V_\mathrm{DA}=sV_\mathrm{F}^\mathrm{IDA}$. As discussed in previous papers \cite{Mennucci2011, Harriman2010,Juzeliunas1994}, the environmental effects can be divided into two contributions $s=f_s^2\cdot1/\epsilon_\mathrm{r}$, changes in effective dipole strengths, $f_s$, and Coulomb screening, $1/\epsilon_\mathrm{r}$. For the first term, the Onsager expression for empty cavity \cite{Aubret2019} is often used\cite{Mennucci2011}, $f_s = 3\epsilon_\mathrm{r}/(2\epsilon_\mathrm{r}+1)$. This gives $s\sim0.55$ for the medium with typical $\epsilon_\mathrm{r}=3$. For simplicity, we also take the orientational factor $\kappa^\mathrm{DA}$ in Eq. (\ref{IDA}) equal to unity.

The estimates of exciton transfer rates between NHC iridium complexes obtained using the Marcus and Levich-Jortner expressions (Eqs. \ref{eq_marcus2} and \ref{eq_levith}, respectively) are presented in Table \ref{table:rates} (the transition dipole moments used in the calculation can be found in Supplementary Material). The values computed with the Marcus and Levich-Jortner expressions are similar qualitatively and in many cases even quantitatively. The data also confirm the qualitative conclusions drawn before through the analysis of the absorption spectra (Fig. \ref{fig:singlet}). 

As expected, because of the huge shift between the emission spectrum of the NHC blue emitter and absorption spectrum of neutral DPBIC, the rate of exciton transfer to DPBIC is negligibly small compared to the transfer to the NHC blue emitter (Table \ref{table:rates}). Although the exact value of this rate is sensitive to the model employed, it is vanishing for both of the models considered.

There is also a significant energy shift between the emission peak and absorption edge of the NHC blue emitter itself. Furthermore, the oscillator strengths of the first optical transitions are rather small and the intermolecular distances between the organometallic complexes are relatively large (exceed 1 nm even for the nearest neighbours) providing small exciton couplings for pairs of emitter complexes as well. As a result, the transfer of excitons between two emitter complexes is also slow (Table \ref{table:rates}). Although the rate increases twice when the Levich-Jortner model is used instead of the Marcus one, the characteristic transfer time in any case is comparable with the previously estimated singlet radiative lifetime ($\tau_\mathrm{r,S} \sim$ 0.1 $\mu$s, see Eq. \ref{eq_RLT}). 

The F\"orster exciton transfer is often characterized using the effective F\"orster radius $R^\mathrm{F}$ (Refs. \onlinecite{Eersel2014, Eersel2016, Mesta2016, Coehoorn2017, Vries2019, Vries2020, Ligthart2021})
\begin{equation} \label{eq_RF}
k(R) = \frac{1}{\tau_\mathrm{r}}\left(\frac{R^\mathrm{F}}{R}\right)^6,
\end{equation}
where $k$ is the transfer rate and $R$ is the distance between the donor and acceptor. Using $\tau_\mathrm{r,S}\sim$ 0.1 $\mu$s, we estimate the F\"orster radius for singlet diffusion to be
$R^\mathrm{F}_\mathrm{diff,S} \sim$ 1 nm. This means that singlet diffusion is basically suppressed. At most some excitons can be transferred to a nearest neighbour blue emitter before the decay.

The rates of exciton transfer to the charged species are two-three orders of magnitude greater compared to the transfer to the neutral NHC blue emitter (Table \ref{table:rates}) due to the significant overlap between the absorption spectra of the charged species with the emission peak of the blue emitter.  As can be estimated from the equation similar to Eq. \ref{eq_RF},
the F\"orster radius for exciton-polaron quenching is $R^\mathrm{F}_\mathrm{e,S}\sim$ 2.4 nm for electrones and $R^\mathrm{F}_\mathrm{h,S}\sim$ 2.7 nm for holes. The exciton quenching should thus occur with a large probability if an electron or a hole gets within this distance from the exciton.
 
 Integrating the exciton transfer rate for distances greater than some minimal distance $R_\mathrm{min}$, the characteristic rates of exciton-polaron quenching can be estimated as 
\begin{equation} \label{eq_EPQ}
k_\mathrm{p}=  k_0 \left(R^\mathrm{F}_\mathrm{p}\right)^6 \int_{R_\mathrm{min}}^{\infty}\frac{4\pi R^2 dR}{R^6} = \frac{4\pi k_0 \left(R^\mathrm{F}_\mathrm{p}\right)^6}{3R_\mathrm{min}^3},
\end{equation}
where $R^\mathrm{F}_\mathrm{p}$ is the F\"orster radius for exciton-polaron quenching. Using $R_\mathrm{min} \sim 1$ nm, which corresponds to the minimal distance between organometallic complexes in the considered emission layer\cite{Paterson2019}, we obtain  $k_\mathrm{h,S}\sim 2\cdot10^{-11}$ cm$^3/$s for holes and $k_\mathrm{e,S}\sim 8\cdot10^{-12}$ cm$^3/$s for electrons. As seen from these results, the transfer rates to the positively charged species are greater than for the negatively charged emitter implying that in the considered OLED, holes are more efficient in exciton quenching as compared to electrons. 

It should be noted that the transfer rates to the positively charged emitter and DPBIC are comparable and are not sensitive to the model employed. The latter is explained by the fact that the transfer is dominated by the optical transitions relatively close to the emission peak of the NHC blue emitter. For  positively charged  DPBIC, the largest contribution to the transfer rate corresponds to the peak at 2.4 eV, while for the positively charged blue emitter, several peaks in the range from 2.0 eV to 2.6 eV contribute in a similar way. The transfer rate to the negatively charged emitter, on the other hand, increases by a factor of 1.5 when the Levich-Jortner expression is used because of the dominant contribution from the optical transitions far from the emission peak of the NHC blue emitter. The most intensive transfer occurs here to the peak at 2.0 eV. We have checked that the estimated exciton transfer rates to charged species weakly depend on their contributions to the reorganization energies used for them if they lie in the range of 0 -- 0.5 eV.

\begin{table*}
    \caption{Rates of exciton transfer from the excited NHC blue emitter (characterized by $\hbar\Omega_{\mathrm{D}}=2.400$ eV and $|\mathbf{d}_{\mathrm{D}}^{\mathrm{T}}|^2=0.1420$ a.u.$^2$) to different species of the emission layer computed\footnote{Only the most relevant transitions are included. Excitation energies and transition dipole moments are obtained by solving the Casida equation without account of spin-orbit coupling for the isolated complexes using the PBE0 functional, def2-TZVP basis set and def2/J auxiliary basis set.} at temperature $T=$300 K for the nearest-neighbour intermolecular distance of $R=1$ nm.}
   \renewcommand{\arraystretch}{1.2}
   \begin{ruledtabular}
        \begin{tabular}{*{11}{c}}
Acceptor &  $\hbar\Omega_{\mathrm{A}}$ 
 &  $|\mathbf{d}_{\mathrm{A}}^{\mathrm{T}}|^2$ & $V_{\mathrm{DA}}$  & $\Delta$ & $J_\mathrm{M}$ & $k_\mathrm{M}$ & $k_\mathrm{M,tot}$ & $J_\mathrm{LH}$  &  $k_\mathrm{LH}$ & $k_\mathrm{LH,tot}$  \\
 & (eV) & (a.u.$^2$) & (meV) & (eV) & (eV$^{-1}$) & (s$^{-1}$) & (s$^{-1}$) & (eV$^{-1}$) & (s$^{-1}$) &     (s$^{-1}$) \\\hline
Blue emitter &  3.156 & 0.484 &  0.581 &  0.757 &  $1.3\cdot10^{-3}$ & $4.3\cdot10^6$ & $5.6\cdot10^6$ &  $2.7\cdot10^{-3}$ & $8.8\cdot10^6$ & $1.1\cdot10^7$ \\
(neutr.) & 3.183 & 0.675 &  0.687 &  0.783  &  $7.9\cdot10^{-4}$ & $3.6\cdot10^6$ & & $1.6\cdot10^{-3}$ &
$7.2\cdot10^6$ &  \\
 &  3.257 & 0.412 & 0.537 & 0.858  & $1.7\cdot10^{-4}$ & $4.6\cdot10^5$ & & $3.2\cdot10^{-4}$ &
$8.6\cdot10^5$ & \\ 
\\
Blue emitter ($-$) & 1.825 & 0.314 & 0.468 & -0.574 & $3.0\cdot10^{-2}$ & $6.2\cdot10^7$ & $1.2\cdot10^9$ & $1.2\cdot10^{-1}$ &  $2.4\cdot10^8$ & $1.8\cdot10^9$ \\
 & 1.910 & 0.197 & 0.371 & -0.490 & $9.4\cdot10^{-2}$ & $1.2\cdot10^8$ & & $2.2\cdot10^{-1}$ & $2.9\cdot10^8$ & \\
 & 2.003 & 0.441 & 0.555 & -0.396 & $2.7\cdot10^{-1}$ & $8.0\cdot10^8$ & & $4.0\cdot10^{-1}$ & $1.2\cdot10^9$ & \\
 & 2.739 & 0.174 & 0.349 & 0.340 & $4.6\cdot10^{-1}$ & $5.4\cdot10^8$ & & $6.3\cdot10^{-1}$ & $7.4\cdot10^8$ & \\
 & 2.786 & 0.347 & 0.492 & 0.386 & $3.0\cdot10^{-1}$ & $6.9\cdot10^8$ & & $4.5\cdot10^{-1}$ & $1.0\cdot10^9$ & \\
\\
Blue emitter (+) & 2.041 & 0.573 & 0.633 & -0.359 & $3.9\cdot10^{-1}$ & $1.5\cdot10^9$ & $2.9\cdot10^9$ & 
$5.0\cdot10^{-1}$ & $1.9\cdot10^9$ & $3.3\cdot10^{9}$ \\
& 2.227 & 0.064 & 0.212 & -0.172 & 1.4 & $5.9\cdot10^8$ &  & 
1.2 & $5.1\cdot10^8$ &  \\
& 2.272 & 0.067 & 0.217 & -0.127 & 1.6 & $7.3\cdot10^8$ &  & 
1.4 & $6.1\cdot10^8$ &  \\
& 2.309 & 0.075 & 0.230 & -0.090 & 1.8 & $9.1\cdot10^8$ &  & 
1.5 & $7.5\cdot10^8$ &  \\
& 2.365 & 0.042 & 0.171 & -0.034 & 2.0 & $5.6\cdot10^8$ &  & 
1.6 & $4.6\cdot10^8$ &  \\
& 2.398 & 0.054 & 0.195 & -0.002 & 2.0 & $7.3\cdot10^8$ &  & 
1.7 & $6.1\cdot10^8$ &  \\
& 2.435 & 0.068 & 0.218 & 0.035 & 2.0 & $9.0\cdot10^8$ &  & 
1.7 & $7.8\cdot10^8$ &  \\
& 2.742 & 0.494 & 0.587 & 0.342 & $4.5\cdot10^{-1}$ & $1.5\cdot10^9$ &  & 
$6.2\cdot10^{-1}$ & $2.1\cdot10^9$ &  \\
& 2.798 & 0.307 & 0.463 & 0.399 & $2.6\cdot10^{-1}$ & $5.4\cdot10^8$ &  & 
$4.0\cdot10^{-1}$ & $8.2\cdot10^8$ &  \\
\\
 DPBIC & 3.648 & 0.317 & 0.471 & 1.249 & $3.2\cdot10^{-11}$ & $6.8\cdot10^{-2}$ & $6.8\cdot10^{-2}$ & $2.3\cdot10^{-12}$ & $4.9\cdot10^{-3}$ & $4.9\cdot10^{-3}$ \\
(neutr.)  &  3.755 & 0.405 & 0.532 & 1.355 & $3.8\cdot10^{-13}$ & $1.0\cdot10^{-3}$ & & $8.3\cdot10^{-15}$ & 
 $2.2\cdot10^{-5}$ &\\
& 3.784 & 0.128 & 0.299 & 1.385 & $1.0\cdot10^{-13}$ & $8.8\cdot10^{-5}$ &  & $1.6\cdot10^{-15}$ & $1.3\cdot10^{-6}$& \\
 & 3.800 & 0.049 & 0.179 & 1.401 & $5.1\cdot10^{-14}$ & $1.6\cdot10^{-5}$ & & $6.3\cdot10^{-16}$ & 
$1.9\cdot10^{-7}$ & \\
\\
DPBIC (+) & 1.997 & 0.342 & 0.489 & -0.403 & $1.7\cdot10^{-1}$ & $3.8\cdot10^8$ & $4.9\cdot10^{9}$  & 
$3.4\cdot10^{-1}$ & $7.8\cdot10^8$ & $4.6\cdot10^{9}$ \\
 & 2.313 & 0.069 & 0.219 & -0.087 & 2.0 & $9.1\cdot10^8$ & & 1.6 & $7.1\cdot10^8$ & \\
 & 2.390 & 0.273 & 0.437 & -0.010 & 2.3 & $4.1\cdot10^{9}$ & & 1.8 & $3.3\cdot10^{9}$ & \\
 & 2.475 & 0.045 & 0.178 & 0.076 & 2.1 & $6.2\cdot10^8$ & & 1.8 & $5.5\cdot10^8$ & \\
 & 2.501 & 0.143 & 0.316 & 0.102 & 1.9 & $1.8\cdot10^9$ & & 1.8 & $1.7\cdot10^9$ & \\
 & 2.578 & 0.046 & 0.179 & 0.178 & 1.4 & $4.2\cdot10^8$ & & 1.5 & $4.5\cdot10^8$ & \\
 & 2.614 & 0.118 & 0.287 & 0.214 & 1.1 & $8.5\cdot10^8$ & & 1.3 & $1.0\cdot10^9$ & \\
 & 2.751 & 0.508 & 0.596 & 0.352 &$3.1\cdot10^{-1}$ & $1.1\cdot10^9$ & & $5.1\cdot10^{-1}$& $1.7\cdot10^8$ & \\
 & 2.773 & 0.420 & 0.542 & 0.374 &$2.4\cdot10^{-1}$ & $6.8\cdot10^8$ & & $4.1\cdot10^{-1}$ & $1.2\cdot10^8$ & \\
\end{tabular}
\end{ruledtabular}
\label{table:rates}
\end{table*}

\subsection{Triplet excitons}
The exciton transfer rates between emitter complexes have been also estimated with account of spin-orbit coupling based on the spectra computed using the Sternheimer equation (Fig. \ref{fig:triplet}, Table \ref{table:triplet_sp}). The rates of transfer processes in which the donor loses a singlet or triplet exciton and the acceptor acquires a singlet or a triplet exciton are presented in Table  \ref{table:triplet}. A three-fold increase in the singlet transfer rates (S$\to$S) as compared to the calculations without account of spin-orbit coupling (S$_0\to$S$_0$) is related to the increase in the transition dipole moments.

It is seen from Table  \ref{table:triplet}  that when the acceptor is excited to the triplet state, the transfer rates are only 2--3 times smaller compared to the case when when the acceptor stays singlet (T$\to$T and S$\to$T versus T$\to$S and S$\to$S, respectively). Although the oscillator strengths of the triplet peaks of the emitter in the ground-state geometry are considerably smaller than those of the singlet peaks (Table \ref{table:triplet_sp}), they are located at lower frequencies and, therefore, the energy shift with respect to the emission peaks of the donor is smaller. For this reason, the contribution of the processes with the triplet exciton on the acceptor is comparable to those with the singlet exciton on the acceptor.

The transfer rates for the initially triplet exciton, on the other hand, are about 30 times smaller than for the singlet exciton (T$\to$T and T$\to$S versus S$\to$T and S$\to$S, respectively). This is because the oscillator strength of the T$^*$ peak corresponding to the triplet exciton for the emitter in the excited-state geometry is an order of magnitude smaller compared to the peak S$^*$ corresponding to the singlet exciton (Table \ref{table:triplet_sp}) and the energy shift between the absorption and emission is greater for the T$^*$ peak. Therefore, transport of triplet excitons is significantly slower than that of singlet excitons. Using the triplet radiative lifetime $\tau_\mathrm{r,T} \sim$ 1 $\mu$s (see Eq. \ref{eq_RLT}), the F\"orster radius estimated for triplet excitons according to Eq. \ref{eq_RF} is $R^\mathrm{F}_\mathrm{diff,S} \sim$ 0.9 nm, even smaller than for singlet excitons. This value is at the lower bound of the range 1 -- 4 nm computed from first principles for diverse iridium-based complexes \cite{Vries2019, Vries2020} and shows that events of exciton diffusion in the considered emission layer should be scarce, in agreement with the kinetic Monte Carlo study \cite{Paterson2019}.

The estimated rates of singlet and triplet exciton transfer to charged species in the emission layer are listed in Table \ref{table:triplet_quench}. It is seen that the transfer rates of triplet excitons to the positively charged NHC blue emitter and DPBIC are an order of magnitude smaller than the similar rates for singlet excitons. This is because of the ratio of the transition dipole moments for singlet and triplet excitons. The difference in the transfer rates of triplet and singlet excitons to the negatively charged emitter is smaller. Here the transfer is dominated by the absorption peak of the negatively charged emitter at 2.0 eV that is located closer in energy to the phosphorescence peak of the neutral emitter than to the fluorescence one. 

Similar to singlet excitons, the rates of transfer of triplet excitons to charged species are three orders of magnitude higher than to the neutral emitter (Table \ref{table:triplet_quench}). The F\"orster radius for triplet-polaron quenching is $R^\mathrm{F}_\mathrm{e,T}\sim$ 2.8 nm for electrons and $R^\mathrm{F}_\mathrm{h,T}\sim$ 3.0 nm for holes, close to typical values considered in kinetic Monte Carlo simulations \cite{Coehoorn2017, Ligthart2021, Mesta2016, Eersel2014, Eersel2016}. The characteristic rates of triplet-polaron quenching are $k_\mathrm{h,T}\sim3\cdot10^{-12}$ cm$^3/$s for holes and $k_\mathrm{e,T}\sim 2\cdot10^{-12}$ cm$^3/$s for electrons. Note that exciton-polaron quenching rates of $10^{-12}-3\cdot 10^{-11}$ cm$^3/$s were measured for OLEDs with iridium complexes  \cite{Reineke2007, Zhang2014, Wehrmeister2015, Erickson2014, Lee2016, Hershey2016} and various organic semiconductors \cite{Murawski2013, Zhang2010}. 

\begin{table*}
    \caption{Rates of exciton transfer from the excited NHC blue emitter to the neutral emitter in the ground state estimated\footnote{Excitation energies and transition dipole moments are obtained by solving the Sternheimer equation for the isolated complexes with account of spin-orbit coupling using the LDA functional with the scissor operator and real-space grid.} for singlet (S) and triplet (T) excitons at temperature $T=$ 300 K for the nearest-neighbour intermolecular distance of $R=1$ nm.}
   \renewcommand{\arraystretch}{1.2}
    \begin{ruledtabular}
        \begin{tabular}{*{10}{c}}
 Transfer & & $V_{\mathrm{DA}}$ &  $\Delta$ & $J_\mathrm{M}$ & $k_\mathrm{M}$ & $k_\mathrm{M,tot}$ &$J_\mathrm{LH}$ & $k_\mathrm{LH}$ & $k_\mathrm{LH,tot}$   \\
 & & (meV) & (eV) & (eV$^{-1}$) & (s$^{-1}$) & (s$^{-1}$) & (eV$^{-1}$) & (s$^{-1}$) & (s$^{-1}$)  \\
 \hline
T$\to$T  & T$^*\to$T$_1$ & 0.012 & 0.708 & $3.3\cdot10^{-3}$ & $4.6\cdot10^3$ & $1.3\cdot10^5$ & $6.8\cdot10^{-3}$ & $9.4\cdot10^3$ & $2.7\cdot10^5$\\
 & T$^*\to$T$_2$ &  0.060  &  0.715  &  $3.0\cdot10^{-3}$  & $1.0\cdot10^5$  & & $6.1\cdot10^{-3}$ &  $2.1\cdot10^5$ &  \\
 & T$^*\to$T$_3$ &  0.030 & 0.737&  $2.0\cdot10^{-3}$ & $1.7\cdot10^4$  & &   $4.0\cdot10^{-3}$ & 
 $3.4\cdot10^4$ & \\
&T$^*\to$T$_4$&  0.016 & 0.742&  $1.8\cdot10^{-3}$  &$4.2\cdot10^3$  & &  $3.6\cdot10^{-3}$  & $8.5\cdot10^3$& \\
& T$^*\to$T$_5$& 0.030 & 0.744 & $1.7\cdot10^{-3}$ & $1.5\cdot10^4$ & & $3.5\cdot10^{-3}$  &  $3.1\cdot10^4$& \\
& T$^*\to$T$_6$ &  0.084  &  0.770 &  $1.0\cdot10^{-3}$ & $6.9\cdot10^4$  & & $2.1\cdot10^{-3}$ & $1.4\cdot10^5$  & \\
& T$^*\to$T$_7$ &  0.062  &  0.777 &  $8.9\cdot10^{-4}$  &$3.3\cdot10^4$ & & $1.8\cdot10^{-3}$ & $6.7\cdot10^4$ & \\
&T$^*\to$T$_8$ & 0.022 & 0.779& $8.6\cdot10^{-4}$ & $4.0\cdot10^3$ & & $1.7\cdot10^{-3}$ & $8.1\cdot10^3$ &\\
\\
T$\to$S & T$^*\to$S$_1$& 0.205 & 0.811 & $4.5\cdot10^{-4}$ & $1.8\cdot10^5$ & $2.9\cdot10^5$ & $8.9\cdot10^{-4}$  & $3.6\cdot10^5$  & $5.8\cdot10^5$ \\
& T$^*\to$S$_2$ & 0.246 & 0.817 & $4.0\cdot10^{-4}$ & $2.3\cdot10^5$ && $7.8\cdot10^{-4}$ & $4.5\cdot10^5$ &  \\
 \\
S$\to$T &  S$^*\to$T$_1$ & 0.034 & 0.646 & $9.8\cdot10^{-3}$ & $1.1\cdot10^5$ & $3.2\cdot10^6$ & $2.0\cdot10^{-2}$ & $2.2\cdot10^5$ & $6.6\cdot10^6$\\
& S$^*\to$T$_2$ &  0.171 & 0.652 & $8.8\cdot10^{-3}$ & $2.5\cdot10^6$ & & $1.8\cdot10^{-2}$ & $5.0\cdot10^6$ &  \\
& S$^*\to$T$_3$ & 0.085 & 0.674 & $6.1\cdot10^{-3}$ & $4.1\cdot10^5$ && $1.2\cdot10^{-2}$ & $8.5\cdot10^5$ &\\
& S$^*\to$T$_4$ & 0.045 & 0.679 & $5.5\cdot10^{-3}$ & $1.0\cdot10^5$ && $1.1\cdot10^{-2}$ & $2.1\cdot10^5$ &\\
& S$^*\to$T$_5$ & 0.086 & 0.682 & $5.3\cdot10^{-3}$& $3.8\cdot10^5$ && $1.1\cdot10^{-2}$ & $7.7\cdot10^5$  & \\
& S$^*\to$T$_6$ & 0.237 & 0.707 & $3.4\cdot10^{-3}$ & $1.8\cdot10^6$ & & $6.9\cdot10^{-3}$ & $3.7\cdot10^6$ & \\
& S$^*\to$T$_7$ & 0.177 & 0.715 & $3.0\cdot10^{-3}$ & $8.8\cdot10^5$ & & $6.1\cdot10^{-3}$ & $1.8\cdot10^6$ & \\
& S$^*\to$T$_8$ & 0.063 & 0.716 & $2.9\cdot10^{-3}$ & $1.1\cdot10^5$ & & $5.9\cdot10^{-3}$ & $2.2\cdot10^5$ & \\
\\
S$\to$S & S$^*\to$S$_1$ & 0.582 & 0.749 & $1.6\cdot10^{-3}$ & $5.0\cdot10^6$ & $8.2\cdot10^6$ & $3.2\cdot10^{-3}$ & $1.0\cdot10^7$ & $1.7\cdot10^7$\\
& S$^*\to$S$_2$ & 0.699 & 0.754 & $1.4\cdot10^{-3}$ & $6.5\cdot10^6$ & & $2.9\cdot10^{-3}$ & $1.3\cdot10^7$ & \\
\\
S$_0\to$S$_0$ & S$^*_0\to$S$_{0,1}$ & 0.401 & 0.757 & $1.3\cdot10^{-3}$ & $2.1\cdot10^6$ & $3.1\cdot10^6$
& $2.7\cdot10^{-3}$ & $4.2\cdot10^6$ & $6.2\cdot10^6$ \\
& S$^*_0\to$S$_{0,2}$ & 0.491 & 0.772 & $9.8\cdot10^{-4}$ & $2.3\cdot10^6$ && $2.0\cdot10^{-3}$ & $4.6\cdot10^6$ & \\
\end{tabular}
 \end{ruledtabular}
\label{table:triplet}
\end{table*}

\begin{table}
    \caption{Rates of transfer of singlet (S) and triplet (T) from the excited NHC blue emitter to charged species estimated\footnote{Excitation energies and transition dipole moments are obtained for the isolated complexes using the Casida equation without account of spin-orbital coupling for the PBE0 functional, def2-TZVP basis set and def2/J auxiliary basis set and Sternheimer equation with account of spin-orbit coupling for the LDA functional with the scissor operator and real-space grid.} at temperature $T=$ 300 K for the nearest-neighbour intermolecular distance of $R=1$ nm.}
   \renewcommand{\arraystretch}{1.2}
    \begin{ruledtabular}
        \begin{tabular}{*{4}{c}}
Acceptor & Transfer &  $k_\mathrm{M,tot}$ (s$^{-1}$)  & $k_\mathrm{LH,tot}$ (s$^{-1}$)   \\\hline
Blue emitter ($-$)  &  T$\to$S  &  $4.0\cdot10^8$  & $4.4\cdot10^8$  \\
 &  S$\to$S  & $2.1\cdot10^9$  &  $2.8\cdot10^9$  \\
\\
Blue emitter ($+$)  &  T$\to$S  &  $7.5\cdot10^8$  &  $7.3\cdot10^8$  \\
 &  S$\to$S   &   $4.7\cdot10^{9}$   &  $5.1\cdot10^{9}$ \\
\\
DPBIC ($+$)  & T$\to$S   & $7.7\cdot10^{8}$   & $7.7\cdot10^{8}$  \\
 & S$\to$S   &  $7.0\cdot10^{9}$  &  $6.5\cdot10^{9}$ \\\hline
\end{tabular}
\end{ruledtabular}
\label{table:triplet_quench}
\end{table}

\section{Conclusions}
Using first-principles calculations, we have investigated exciton transfer between NHC iridium (III) complexes that can play the roles of emitters as well as electron and hole transporters in OLEDs. Similarities in the geometry and electronic structure evolution upon the excitation have been found for DPBIC and structurally close NHC blue emitter. In both of the complexes, the excitation is accompanied by
the significant reorganization energy of about 0.7 eV and considerable loss in the oscillator strength. The latter is related to the strong metal-ligand charge transfer character of the first optical transitions. The high reorganization energy and small oscillator strength for the emission peak result in low exciton diffusion. 

The analysis of the absorption spectra for charged complexes in the emission layer shows that within 0.7 eV from the emission peak of the blue emitter there are a number of optical transitions in charged species, which can lead to exciton quenching. Some of these transitions in the negatively charged emitter complex resemble the first optical transitions in the neutral emitter. In the positively charged species, these are transitions from low-energy states to the state coming from HOMO of the neutral complex.

Singlet-triplet splittings of 0.06 eV and 0.1 eV are found for the NHC blue emitter in the excited and ground states via calculations with account of spin-orbit coupling. The oscillator strengths for the triplet excitations are found to be an order of magnitude smaller than for the singlet ones.

The real-time propagation for a dimer consisting of two emitter complexes in the excited and ground state geometry separated by 1.2 nm reveals only small changes in the spectra of the complexes related to the presence of each other. Calculations of the exciton coupling via explicit integration of transition densities shows that the Dexter mechanism has a negligibly small contribution to the exciton transfer at such distances compared to the F\"orster mechanism. At the same time, the exciton coupling computed in the ideal dipole approximation, although is correct in the order of magnitude, deviates considerably from the result obtained with account of the spatial distribution of the transition density. This means that accurate simulations of exciton transport in the emission layer require explicit calculations of exciton couplings with account of spatial distributions of transition densities for different arrangements of the donor and acceptor molecules in the emission layer. Only qualitative conclusions on the relative probability of different exciton transfer events can be drawn based on the use of the ideal dipole approximation.

The qualitative analysis of transfer rates to different complexes of the emission layer shows that exciton diffusion is strongly suppressed because of the large difference in emission frequency of the emitter and absorption frequencies of the neutral emitter and DPBIC. On the other hand, rates of exciton transfer to charged species are high, which should lead to significant exciton-polaron quenching. In the considered emission layer, the transfer rates are greater for positively charged species as compared to the negatively charged ones. Recently it has been proposed that exciton quenching can be reduced by combining phosphorescent and fluorescent emitters in a sensitizing approach which allows to reduce the radiative lifetime \cite{Heimel2018, Kim2022, Yan2023}. This and other\cite{Zhao2023} ways to further increase the efficiency of such devices require further investigation. 

\section*{Supplementary Material}
The Supplementary Material includes the comparison of absorption spectra computed for the NHC blue emitter using different exchange-correlation functionals, molecular orbitals of the neutral, charged and excited NHC blue emitter and DPBIC and transition dipole moments for singlet excitations of these complexes.

\begin{acknowledgments}
We thank the financial support from the EU-H2020 project  ``MOSTOPHOS" (n. 646259). JJS gratefully acknowledges the funding from the European Union Horizon 2020 research and innovation program under the Marie Sklodowska-Curie Grant Agreement No. 795246-StrongLights. The authors gratefully thank Prof. Angel Rubio for his comments and support.
\end{acknowledgments}

\section*{Data Availability Statement}
The data that support the findings of this study and the relevant code are openly available in the Zenodo repository at \url{https://doi.org/10.5281/zenodo.10605154}, reference number 10605154.\cite{Lebedeva2024}

\bibliography{oled_refs}

\end{document}